\documentclass[%
 reprint,
 amsmath,amssymb,
 aps,
]{revtex4-1}

\usepackage{color}

\usepackage{hyperref}
\usepackage{graphicx}
\usepackage{dcolumn}
\usepackage{bm}
\usepackage{color}

\newcommand{\beginsupplement}{%
        \setcounter{table}{0}
        \renewcommand{\thetable}{S\arabic{table}}%
        \setcounter{figure}{0}
        \renewcommand{\thefigure}{S\arabic{figure}}%
}

\begin{document}

\preprint{APS/123-QED}

\title{The modular organization of human anatomical brain networks:\\ Accounting for the cost of wiring}

\author{Richard F. Betzel$^1$}
\author{John D. Medaglia$^{1,2}$}
\author{Lia Papadopoulos$^3$}
\author{Graham Baum$^4$}
\author{Ruben Gur$^4$}
\author{Raquel Gur$^4$}
\author{David Roalf$^4$}
\author{Theodore D. Satterthwaite$^4$}
\author{Danielle S. Bassett$^{1,5}$}
 \email{dsb @ seas.upenn.edu}
\affiliation{
 $^1$Department of Bioengineering, University of Pennsylvania, Philadelphia, PA, 19104
}
\affiliation{
 $^2$Department of Psychology, University of Pennsylvania, Philadelphia, PA, 19104
}
\affiliation{
	$^3$Department of Physics, University of Pennsylvania, Philadelphia, PA, 19104
}
\affiliation{
$^4$Neuropsychiatry Section, Department of Psychiatry, University of Pennsylvania, Philadelphia, PA, 19104
}
\affiliation{
$^5$Department of Electrical and Systems Engineering, University of Pennsylvania, Philadelphia, PA, 19104
}

\date{\today}
\begin{abstract}
Brain networks are expected to be modular. However, existing techniques for estimating a network's modules make it difficult to assess the influence of organizational principles such as wiring cost reduction on the detected modules. Here, we present a modification of an existing module detection algorithm that allows us to focus on connections that are unexpected under a cost-reduction wiring rule and to identify modules from among these connections. We apply this technique to anatomical brain networks and show that the modules we detect differ from those detected using the standard technique. We demonstrate that these novel modules are spatially distributed, exhibit unique functional fingerprints, and overlap considerably with rich clubs, giving rise to an alternative and complementary interpretation of the functional roles of specific brain regions. Finally, we demonstrate that, using the modified module detection approach, we can detect modules in a developmental dataset that track normative patterns of maturation. Collectively, these findings support the hypothesis that brain networks are composed of modules and provide additional insight into the function of those modules.
\end{abstract}

\maketitle
\section*{Introduction}

Modular organization is a hallmark of complex networks. This means that a network's nodes can be partitioned into internally dense and externally sparse subnetworks called modules or communities \cite{porter2009communities, newman2012communities}. This type of organization has been observed in biological neural networks at virtually all spatial scales \cite{meunier2010modular, sporns2016modular}, from cellular networks of synaptically-coupled neurons \cite{jarrell2012connectome, shimono2015functional, lee2016anatomy} up to whole-brain networks of regions linked by white-matter fiber tracts \cite{hagmann2008mapping, bassett2011conserved, betzel2013multi, lohse2014resolving}.

%\textcolor{blue}{The cost of a brain's wiring as measured by its total length of connections and energy consumption constrains its possible network architectures \cite{laughlin2003communication, kaiser2006nonoptimal, henderson2011geometric, henderson2013using, chen2006wiring, avena2014using}. The architecture of brain networks, in turn, influences brain function (e.g. by shaping the pattern of functional interactions at rest and during tasks \cite{honey2009predicting, hermundstad2013structural,honey2009predicting,mivsic2015cooperative}), as evidenced by the growing number of studies that have reported associations between different network properties and behavior. Therefore, our understanding of how brain networks are related to cognition and behavior hinges, in part, on our ability to understand how they are shaped and constrained by their wiring cost.}

Why do biological neural networks tend to be modular? One parsimonious explanation is that having modules generally leads to networks that are more fit than those without modules \cite{gerhart2007theory}. This improved fitness is the result of a confluence of factors. For example, modular networks can engage in specialized information processing \cite{espinosa2010specialization}, perform focal functions \cite{baldassano2016topological}, and support complex neural dynamics \cite{gallos2012small}. The near-autonomy of modules also means that they can be interchanged or modified without influencing the rest of the system, thereby enhancing the network's robustness, phenotypic variation, and evolvability -- the system's capacity to explore novel adaptive configurations \cite{kirschner1998evolvability}. In addition, modules serve as buffers of deleterious perturbations to the network -- an insult will remain confined to the module where it originated rather than spreading across the network \cite{nematzadeh2014optimal}. Finally, modularity allows for an efficient embedding of a network in physical space such as the three-dimensional space of the skull \cite{bassett2010efficient}.

Another organizational principle that contributes to the brain's modular organization, and indeed to its network architecture more generally, is its apparent drive to reduce its cost of wiring \cite{laughlin2003communication,chen2006wiring,raj2011wiring}. The formation and maintenance of fiber tracts requires material and energy, resources that the brain possesses in limited quantity and therefore must allocate judiciously \cite{bullmore2012economy}. This economy of resources results in a distribution of connection lengths skewed in favor of short, low-cost connections \cite{samu2014influence,roberts2016contribution, henderson2011geometric, henderson2013using}.

While brain networks clearly favor short-range connections, the brain does not minimize its wiring cost in a strict sense and allows for the formation of a small number of long-distance connections. These costly connections are, by definition, inconsistent with the hypothesis that brain network architecture is optimized according to a cost-minimization principle \cite{ercsey2013predictive, song2014spatial}. Instead, they are the result of a trade-off between the formation of connections that reduce the network's wiring cost and those that improve its functionality. We argue, here, that shifting focus onto these long, costly connections can be useful in facilitating a deeper understanding of the brain's modular structure and its function. Our argument is based on two observations.

First, long-distance connections are particularly important for brain function. In principle, costly, long-distance connections could have been eliminated over the course of evolution if the brain were strictly optimized to minimize its wiring cost \cite{van2016comparative}. The existence of such connections, however, implies that they improve brain network fitness more so than had they been replaced by shorter, less-costly connections. We speculate that this additional fitness is a direct result of specific functional advantages that long-distance connections confer to neural systems. For example, long connections improve the efficacy of interregional communication and information transfer by reducing the average number of processing steps between neural elements \cite{kaiser2006nonoptimal, bassett2006small} and by linking high-degree hub regions together to form integrative cores \cite{hagmann2008mapping} and rich clubs \cite{van2011rich,van2012high}. Less is known, however, about the modular organization of the brain's long-distance architecture. Shifting emphasis onto longer connections will allow to uncover such modules, should they exist, and enhance our understanding of their functional roles.

Second, our primary tools for detecting brain network modules are biased by the presence of short-range connections, and by shifting emphasis onto long-range connections we can mitigate the effects of this bias. Because brain network's are large and their wiring patterns complicated, we usually cannot identify modules simply from a visual inspection of the network. Rather, we rely on \emph{module detection} tools to uncover modules algorithmically \cite{fortunato2010community, palla2005uncovering, rosvall2008maps, ahn2010link, lancichinetti2011finding, peixoto2013parsimonious}. Of these techniques, the most popular is centered around a quality function known as \emph{modularity} (or simply $Q$) \cite{newman2004finding}. Modularity measures the quality of a nodal partition as the difference between the observed number of within-module connections and the number of such connections expected under some null model \cite{newman2004finding}. Greater modularity values are taken to indicate higher quality partitions, and the partition that maximizes modularity is treated as a reasonable estimate of a network's modular organization.
	
\begin{figure*}[t]
	\begin{center}
		\centerline{\includegraphics[width=1\textwidth]{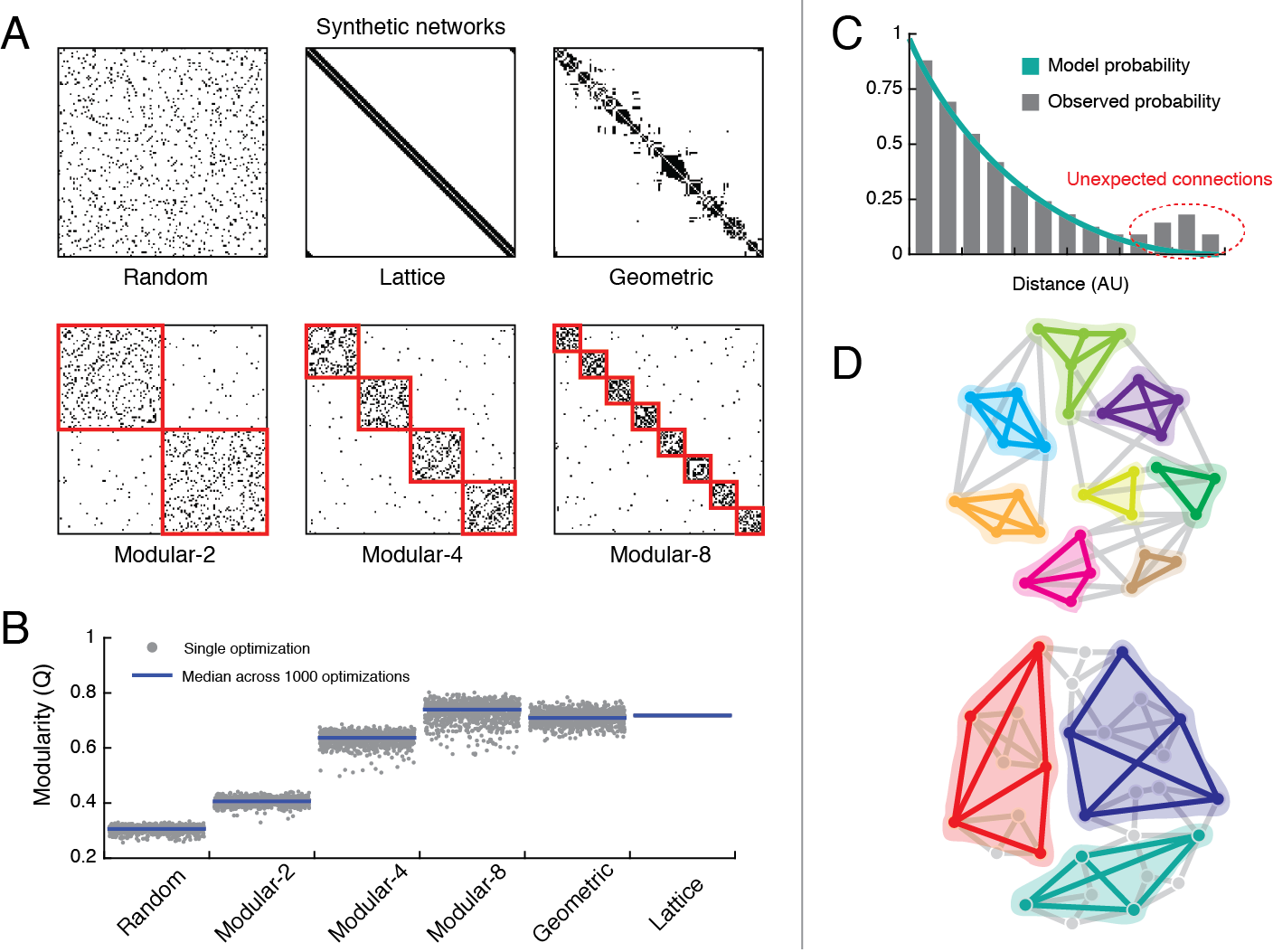}}
		\caption{\textbf{Synthetic networks and an illustration of the problem.} (\emph{A}) We show six synthetic networks with the same connection density, three of which are amodular (random, lattice, and geometric); the remaining three have two, four, and eight modules respectively. (\emph{B}) The modularity function, $Q$, is greatest for the eight-module network, but the lattice and geometric networks, though formed through fundamentally amodular generative processes, exhibit the next-greatest $Q$ values. This indicates that $Q$ can mistakenly give the impression that networks with no modules are, in fact, highly modular. (\emph{C}) The majority of connections in anatomical brain networks are short-range and can be accounted for parsimoniously by a cost-reduction mechanism. Our aim is to perform module detection on observed connections that are unanticipated by a cost-reduction mechanism; these connections tend to be long-distance connections, as they are more costly. (\emph{D}) The result of this refocusing is that, instead of modules whose internal connections are short-range (\emph{top}), we detect modules linked by long-distance connections (\emph{bottom}).} \label{modulesToy}
	\end{center}
\end{figure*}

Oftentimes, we use the modularity score, itself, to assess whether an observed network is or is not modular. This involves comparing its modularity with that of an appropriately constructed random network, which cannot be partitioned into meaningful modules and is therefore associated with low modularity \cite{maslov2002specificity}. If the observed modularity is statistically greater than that of a random network ensemble, then we have evidence that the network is modular \cite{guimera2005functional, reichardt2006statistical}. In random geometric networks, however, the formation of connections depends only on the distance between two nodes \cite{dall2002random} (Fig.~\ref{modulesToy}A,B). Though formed through a fundamentally amodular generative process, these networks are associated with greater than expected modularity, and based on the aforementioned criterion, would be misclassified as modular. This indicates that the modularity of networks with strong spatial constraints or local clustering (e.g. lattice networks) can be misinterpreted as evidence that the network is, in fact, modular \cite{karrer2008robustness}.

This presents a problem when we perform module detection on biological neural networks, for which possible cost-reduction principles have led to an over-representation of short-range connections. Can we be sure that the modules we uncover are not merely the effect of spatial constraints? One possible strategy for mitigating this concern is to discount all elements of the network that are consistent with a spatial wiring rule and search for modules among the residual elements -- i.e. long connections. Such a strategy, incidentally, could be realized under the modularity maximization framework by redefining the modularity equation and replacing the standard null model with one based on a spatial wiring rule \cite{expert2011uncovering}. This redefinition results in the detection of modules whose internal density of connections exceeds what would be expected had the network been generated strictly based on a spatial wiring rule (Fig.~\ref{modulesToy}C,D). This modification is in the same spirit as past studies in which the modularity of spatially-wired networks was compared to observed brain networks \cite{samu2014influence,roberts2016contribution, henderson2011geometric, henderson2013using, betzel2016generative}.

The rest of this report describes a theoretical framework for drawing focus to long-distance connections and studying their modular organization. We develop a spatial null model for structural brain networks, which we integrate into the modularity maximization framework. This seemingly small modification allows us to detect novel modules, which we show are consistent across individuals and have unique functional fingerprints. The modules we detect also suggest alternative functional roles for specific brain regions and systems. In particular, we find that somatosensory cortex appears as an integrative structure whereas attentional, control, default mode, and visual systems now appear more segregated from the rest of the brain. Additionally, we investigate the relationship of these modules with the brain's rich club. Whereas traditional rich club analysis suggests that rich club regions are distributed across modules, we show that rich club regions tend to cluster within the same modules. Finally, we apply our approach to a developmental dataset and show that, among the modules we detect, one in particular appears to track with developmental age. This final component suggests that this framework for module detection is not only a methodological advance, but also a practical and sensitive tool to address specific neuroscientific hypotheses. Ultimately, the framework proposed here offers a novel perspective on the brain's modular organization and serves to complement our current understanding of brain network function.

\section*{Materials and Methods}

\begin{figure*}[t]
	\begin{center}
		\centerline{\includegraphics[width=1\textwidth]{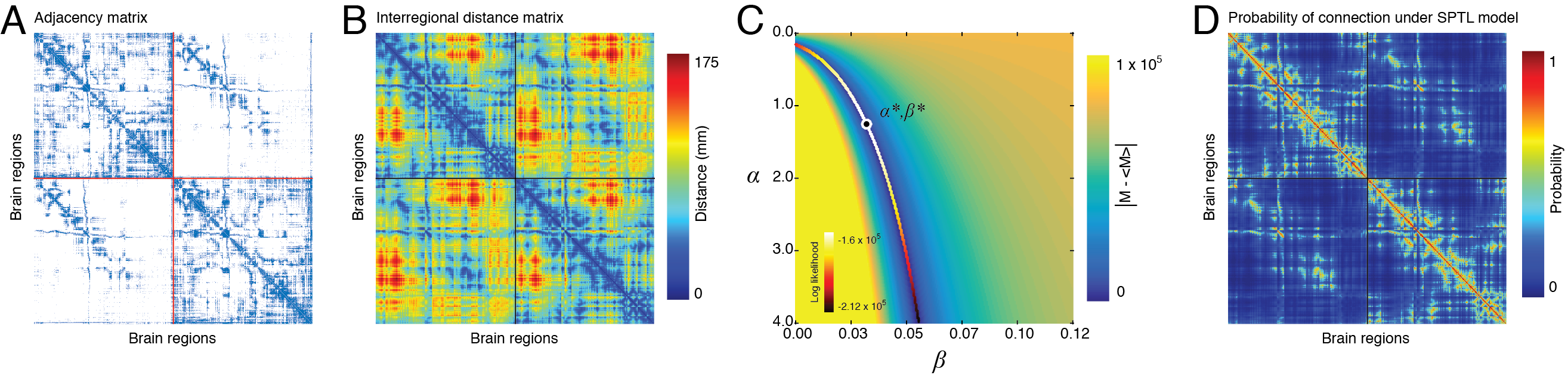}}
		\caption{\textbf{Typical input matrices} (\emph{A}) Representative connectivity matrix for the DSI dataset. (\emph{B}) Interregional distance matrix, calculated as the Euclidean distance between the centroids of the $N$ brain regions (nodes). (\emph{C}) To fit the SPTL model to the observed connectivity matrix, we find the curve through a two-dimensional parameter space (characterized by a density-penalty $\alpha$ and a length-penalty $\beta$) for which the observed number of connections, $M$, is equal to the expected number of connections, $\langle M \rangle$. Along this curve, we then identify the $\alpha^*, \beta^*$ that maximize, $\mathcal{L}$, the log-likelihood that the SPTL model generated the observed connectivity network. (\emph{D}) Fitting the SPTL model returns a matrix whose elements give the probability that any pair of nodes will be connected.} \label{matrices}
	\end{center}
\end{figure*}

\subsection*{Datasets}
We analyzed two human anatomical network datasets: (\emph{1}) a healthy adult cohort constructed from diffusion spectrum imaging data (DSI) and (\emph{2}) a developmental cohort constructed from diffusion tensor imaging (DTI) data.  In the following section we describe, briefly, the strategies used to process these data and to obtain estimates of their modular organization.

\subsubsection*{Human DSI}
The first dataset we analyzed was generated from DSI in conjunction with state-of-the-art tractography algorithms to reconstruct large-scale interregional white-matter pathways for 30 healthy adult individuals. Study procedures were approved by the Institutional Review Board of the University of Pennsylvania, and all participants provided informed consent in writing. Details of the acquisition and reconstruction have been described elsewhere \cite{betzel2016optimally}. We studied a division of the brain into $N=1014$ regions (nodes) \cite{cammoun2012mapping}. Based on this division, we constructed for each individual an undirected and binary connectivity matrix, $\mathbf{A} \in \mathbb{R}^{N \times N}$, whose element $A_{ij}=1$ if at least one streamline (reconstructed fiber tract) was detected between regions $i$ and $j$; otherwise $A_{ij}=0$ (Fig.~\ref{matrices}A). Additionally, we extracted the location of the center of mass for each brain region. From these coordinates, we calculated the Euclidean distance matrix, $\mathbf{D} \in \mathbb{R}^{N \times N}$, whose element $D_{ij}$ gave the distance between regions $i$ and $j$ (Fig.~\ref{matrices}B).

%\textcolor{blue}{Of the 1014 regions defined here, 1000 were located in the cerebral cortex and were selected to have approximately the same volume. The remaining 14 regions corresponded to sub-cortical structures and had larger volumes, which could artificially inflate their degrees. While we analyze the full network (cortical + subcortical) in the main text, we also repeated our module detection on a network composed of only 1000 cortical regions.}

\subsubsection*{Human developmental DTI}
The human DTI data was taken from the Philadelphia Neurodevelopmental Cohort (PNC). Data were acquired in a collaboration between the Center for Applied Genomics at the Children’s Hospital of Philadelphia and the Brain Behavior Laboratory at the University of Pennsylvania. Study procedures were approved by the Institutional Review Board of both institutions. Adult participants provided informed consent; minors provided assent and their parent or guardian provided informed consent. Diffusion data processing and tractography was performed using the same pipeline as the human DSI data, resulting in anatomical brain networks for 1110 individuals aged 8--22 years \cite{satterthwaite2014neuroimaging, tang2016structural}. To ensure high-quality, artifact-free data, we employed a strict exclusion policy \cite{roalf2016impact}. Of the original 1110 individuals, we excluded individuals whose total number of binary connections was beyond $\pm 2$ standard deviations from the group mean. We also excluded subjects with high levels of motion (displacement $>0.5 \text{ mm}$) and poor signal to noise ratio (SNR $<6$) \cite{tang2016structural}. These procedures identified a total of 751 subjects eligible for subsequent analysis. Note, that we did not exclude subjects on the basis of health or medical condition. We parcellated the brain into $N=233$ regions \cite{cammoun2012mapping}. As in the DSI data, regions were considered connected if they were linked by at least one streamline.

\subsubsection*{Group-representative networks}
Within each dataset we pooled network data across individuals to form representative networks. For the DSI dataset we included all 30 individuals and for the DTI dataset we included only adult subjects aged 18--22 years. The common procedure for constructing representative networks involves retaining the connections that are most consistently expressed across individuals; because tractography algorithms are biased towards detecting short connections, these procedures may result in a ``representative'' network with more short-range and fewer long-range connections than is characteristic of any individual subject \cite{roberts2016consistency}. Here, we constructed the representative network so as to (\emph{i}) match the average binary density of subject-level networks while (\emph{ii}) simultaneously approximating the typical edge length distribution. The second step in this procedure was critical, as it ensured that the representative network included the same proportions of short and long connections as the typical individual. Our algorithm for constructing representative networks (an earlier version of which has been described elsewhere \cite{mivsic2015cooperative}) involved, first, estimating the cumulative edge length distribution across all subjects. Next, we sampled $M + 1$ linearly-spaced points along this distribution, where $M$ was the average number of connections exhibited across subjects. Within each percentile bin, we then identified the most consistently detected edge and retained that edge in our representative connectivity matrix. We performed this procedure separately for within- and between-hemispheric connections. Conceptually, this procedure selected the most consistent edges within a given distance range, ensuring that we sampled consistenly-detected short and long connections. In subsequent sections, we show that the modules detected using the representative matrices described here were also consistently expressed at the level of individual subjects.

\subsection*{Modularity maximization}
The principal aim of this report was to modify existing module detection techniques to make them more sensitive to long-distance connections and modules whose emergence cannot be attributed solely to cost-reduction or purely geometry-driven principles. We focused on \emph{modularity maximization}, which is among the most widely-used module detection algorithms in network science \cite{newman2004finding, porter2009communities, fortunato2010community}. The aim of modularity maximization is simple: to partition a network of $N$ nodes into $K$ non-overlapping modules so as to maximize the \emph{modularity quality function}, which measures the difference between the observed number of within-module connections and the number of such connections expected under some null model. If $A_{ij}$ and $P_{ij}$, respectively, are the observed and expected number of connections between nodes $i$ and $j$, then the modularity, $Q$, is calculated as:
  
\begin{equation}
Q = \sum_{ij} [A_{ij} - \gamma P_{ij}] \delta (c_i , c_j).
\end{equation}

\noindent Here we use the variable $c_i \in \{ 1 , \ldots , K \}$, to indicate the module to which node $i$ is assigned. The Kronecker delta function, $\delta (c_i , c_j)$, is equal to unity when $c_i = c_j$ and is zero otherwise. We also include the resolution parameter, $\gamma \in [0, \infty]$. This parameter can be tuned to smaller or larger values so as to detect correspondingly larger or smaller modules \cite{reichardt2006statistical, arenas2008analysis}. 

The process of maximizing $Q$, however, is computationally intractable for all but the most trivial cases. Therefore, to approximate the optimal $Q$ we rely on heuristics, the most widely used being the Louvain algorithm \cite{blondel2008fast}. The Louvain algorithm is a greedy method that is both computationally efficient and performs well in benchmark tests \cite{lancichinetti2009community}. However, it also features a stochastic element, meaning that its output can vary from run to run and should therefore be repeated multiple times \cite{bassett2011dynamic}.

We applied modularity maximization to the representative DSI and DTI connectivity matrices. In both cases, we had no prior knowledge of how to choose the resolution parameter, so we varied $\gamma$ over the interval $[0,5]$ in increments of 0.1 giving us a total of 51 parameter values at which we sought a partition of network nodes into modules. At each such value, we repeated the Louvain algorithm 500 times. We also repeated this module detection procedure for each of the 30 individuals in the DSI dataset. Due to the prohibitively large number of participants, we did not perform individual-level modularity maximization for the participants in the PNC cohort (DTI dataset).

\subsubsection*{Selecting the resolution parameter}
Modularity maximization resulted in 500 estimates of network modules at each of the 51 resolution parameter values. Which of these parameters should we focus on? Which estimate of the network's modules should we believe? In this section we justify and explain our approach to answering these questions.

First we note that there is no definitive rule for choosing $\gamma$. One possible heuristic, however, is to identify the parameter at which modules are especially well-defined \cite{bassett2013robust}. Intuitively, if the modules are well-defined, then they are also easily detectable \cite{chai2016functional}. Therefore, we focused on the resolution parameter where repeated runs of the Louvain algorithm resulted in similar module estimates \cite{doron2012dynamic}. The procedure for identifying such values entailed, at each value of $\gamma$, calculating the average similarity over all pairs of detected partitions and focusing on the $\gamma$ at which the average similarity was greatest. As a measure of similarity, we used the \emph{z}-score of the Rand coefficient \cite{traud2011comparing}. For two partitions, $X$ and $Y$, we measure their similarity as:

\begin{equation}
Z_{XY} = \frac{1}{\sigma_{w_{XY}}}w_{XY} - \frac{W_X W_Y}{W}.
\end{equation}

\noindent Here, $W$ is the total number of node pairs in the network, $W_X$ and $W_Y$ are the number of pairs in the same modules in partitions $X$ and $Y$, respectively, $w_{XY}$ is the number of pairs assigned to the same module in \emph{both} $X$ and $Y$, and $\sigma_{w_{XY}}$ is the standard deviation of $w_{XY}$. The value of $Z_{XY}$ can be interpreted as how great, beyond chance, is the similarity of partitions $X$ and $Y$.

\subsubsection*{Constructing consensus modules}

The above procedure allowed us to isolate a single resolution parameter and corresponding partition ensemble for subsequent analysis. However, the partition ensemble may contain dissimilar partitions \cite{good2010performance}. To resolve this variability we constructed a consensus partition that summarized the commonalities of partitions within the ensemble \cite{lancichinetti2012consensus, bassett2013robust}. To construct such a partition we employed an association-reclustering framework. This procedure involved two main steps. The first step involved computing the association matrix, $\mathbf{T} \in \mathbb{R}^{N \times N}$, from the partition ensemble. The matrix element $T_{ij}$ was equal to the number of times that nodes $i$ and $j$ were co-assigned to the same module. The association matrix can be thought of as encoding the strength of modular relationships between pairs of nodes. The second step involved reclustering the association matrix using modularity maximization to identify consensus modules. We defined the consensus modularity function as:

\begin{equation}
Q^{\text{CONS}} = \sum_{ij} [T_{ij} - \langle T_{ij} \rangle] \delta (c_i^{\text{CONS}},c_j^{\text{CONS}}).
	\end{equation}
 
\noindent Here, $c_i^{\text{CONS}}$ represents an estimate of the consensus module assignment for node $i$. The variable, $\langle T_{ij} \rangle$, is the expected number of times that nodes $i$ and $j$ would be co-assigned to the same module if the module assignments were randomly permuted. This value can be calculated exactly from the matrix, $\mathbf{T}$, as $\langle T_{ij} \rangle = \frac{2}{N(N-1)} \sum_{i,j>i} T_{ij}$.

We found that maximizing $Q^{CONS}$ yielded partitions that were more consistent with one another than those that made up the partition ensemble. If repeated maximization of $Q^{CONS}$ yielded identical partitions, then we considered any one of those partitions to be a good estimate of the consensus partition and the association-reclustering algorithm terminated. If, after many repetitions there was still unresolved variability, we constructed from the estimates of $c_i^\text{CONS}$ a new association matrix and repeated the algorithm. In practice, we found that the algorithm converged in two or fewer iterations. The consensus clustering approach allowed us to obtain from an ensemble of partitions a single consensus partition for each participant at each $\gamma$ value.

\subsubsection*{Statistical significance of modules}
Modularity maximization will always partition a network into modules, even when the network has no true modules \cite{guimera2004modularity}. It is good practice to test the statistical significance of modules by comparing them against a null model. Here, we tested the statistical significance by calculating the modularity contribution of each module, $c$:

\begin{equation}
Q_c = \sum_{ij \in c} [A_{ij} - \gamma P_{ij}],
\end{equation}

\noindent which we compared against a null model wherein we permute module assignments uniformly at random (10000 times) while preserving the total number and size of modules. For a module to be considered statistically significant, its modularity contribution had to exceed the 99$^{\mathrm{th}}$ percentile of the null model.

\subsection*{Null models}
In the modularity equation, the term $P_{ij}$ represents the expected number of connections between nodes $i$ and $j$ given some null connectivity model. Throughout the previous sections, we left this term undefined.  The precise value of $P_{ij}$, however, depends on the nature of the null model selected by the user. The most common choice is the Newman-Girvan (NG) model \cite{porter2009communities}. The NG model generates synthetic networks with the precise degree sequence observed in the real network, but where connections are otherwise made uniformly at random. Under this model, the expected number of connections between nodes $i$ and $j$ is given by:

\begin{equation}
P_{ij}^{\text{NG}} = \frac{k_i k_j}{2m},
\end{equation}

\noindent where, $k_i = \sum_j A_{ij}$, is the degree of node $i$ and $2m = \sum_i k_i$ is the total number of connections in the network.

The NG model tests the hypothesis that an observed network's modules are a consequence of its degree sequence. However, other null models can be used to test other hypotheses \cite{bassett2015extraction,papadopoulos2016evolution}. In this report, we wished to test whether a network's modules were a consequence of a cost-reduction wiring rule. To do so, we needed a cost-reduction null model.

Cost-reduction can be viewed as a preference for shorter, and hence less-costly, connections, suggesting that a network's spatial embedding is critical for determining its cost \cite{barthelemy2011spatial}. Under a cost-reduction wiring rule, then, the probability of forming a connection between two nodes should decay monotonically as a function of distance. To match this intuition, we propose the following spatial model \cite{kaiser2004spatial}:

\begin{equation}
P_{ij}^{\text{SPTL}} = \min(1, \alpha e^{-\beta D_{ij}})
\end{equation}

\noindent where, $D_{ij}$ is the Euclidean distance separating nodes $i$ and $j$. The free parameters, $\{ \alpha , \beta \} \in [0, \infty \}$, control the overall likelihood of forming connections and the extent to which connections are penalized for their length, respectively. Note that here we use the Euclidean (straight line) distance to measure the cost of forming a connection between two brain regions. A more accurate measure of a connection's cost would take into account its curvilinear trajectory through space -- its fiber length. However, because we only have fiber length estimates for connections detected by the tractography algorithm and because the cost-reduction model considers all connections and not only those that were detected, we used Euclidean distance as a proxy for fiber length. We confirmed that, for existing connections, these two measures are highly correlated, suggesting that Euclidean distance may be an acceptable approximation of fiber length for our purposes ($r = 0.696$, $p < 10^{-15}$; Fig.~S6)

\subsubsection*{Fitting the spatial model}
The spatial null model featured two free parameters: the density-penalty $\alpha$ and the length-penalty $\beta$. We selected these parameters using a simple two-step procedure. First, we sampled 1001 linearly-spaced values over the range $\alpha \in [0, \alpha_{max}]$. For both DSI and DTI data, we set $\alpha_{max} = 10$. For each value of $\alpha$, we used the bisection method to find the $\beta$ value corresponding to the spatial model whose number of expected edges, $\langle M \rangle$, is equal to $M$, the observed number of edges \cite{burden1985numerical}. This procedure resulted in a curve through parameter space where any $\{ \alpha, \beta \}$ along the curve satisfied $\langle M \rangle = M$ (Fig.~\ref{matrices}C). For each such pair, we calculated the log-likelihood that the spatial model, given those parameters, generated the observed network:

\begin{equation}
\mathcal{L} = \sum_{ij} log [P_{ij}^{A_{ij}}(1 - P_{ij})^{1-A_{ij}}],
\end{equation}

\noindent where, $P_{ij} = P_{ij}^{\text{SPTL}}$. We subsequently focused on $\{ \alpha^* , \beta^* \}$, the pair of parameters that maximized $\mathcal{L}$ (Fig.~\ref{matrices}D). Thus, the $P_{ij}^{\text{SPTL}}$ that we focused on corresponded to the null model constrained to have, on average, the same number of connections as the observed network and, from among that subset of models, was the one most likely to have generated the observed brain network. It should be noted that rather than enforcing the model to have the same number of connections as the observed network, we could have selected an alternative measure -- e.g. total wiring cost. Our decision to focus on models with the same number of connections as the observed network is in line with the standard practices in the field, wherein networks are compared against null models with the same binary density \cite{van2010comparing}.

\subsubsection*{Modularity maximization pipeline summary}
In summary, our analysis pipeline took as input a connectivity matrix and the three-dimensional locations of each network node. We calculated, under the Newman-Girvan and a spatial null model, the expected number of connections between all pairs of nodes. We compared these values to those estimated in the observed network, which (along with a resolution parameter) allowed us to define two separate modularity functions: one using the NG null model and another using the SPTL model. We optimized these modularities using the Louvain algorithm, identified an optimal resolution parameter, estimated consensus modules, and calculated each module's statistical significance.

\subsection*{Network statistics}
The previous sections were devoted to the enterprise of modularity maximization for module detection, which is the focus of this report. Elsewhere in our analysis, we computed other metrics, either directly on a network or based on detected modules. In this section, we define those metrics.

\subsubsection*{Participation coefficient}
Given a partition of a network's nodes into modules, one can calculate each node's participation coefficient, which describes how its connections are distributed across modules \cite{guimera2005functional}. The participation coefficient of node $i$ is calculated as:

\begin{equation}
P_i = 1 - \sum_c  \bigg( \frac{ \kappa_{ic}}{k_i} \bigg)^2.
\end{equation}

\noindent Here, $\kappa_{ic}$ is the number of connections node $i$ makes to module $c$ and $k_i$ is the degree of node $i$. A value of $P_i$ close to one indicates that a node's connections are uniformly distributed over modules, while a value close to zero indicates that the majority of a node's connections are made to its own module.

\subsubsection*{Rich club detection}
A rich club is a collection of high-degree nodes that are more interconnected to one another than expected by chance \cite{colizza2006detecting}. We denote the set of nodes that make up a rich club as $r$. Rich clubs are detected by calculating the rich club coefficient:

\begin{equation}
\phi (k) = \frac{2 E_{>k}}{N_{>k}(N_{>k} - 1)},
\end{equation}

\noindent which gives the density of connections between nodes of degree greater than $k$. This coefficient is then compared against chance, where chance is a degree-preserving null model where connections are otherwise formed at random \cite{maslov2002specificity}. The rich club coefficient is then typically expressed as a normalized rich-club coefficient -- the observed coefficient divided by the mean across an ensemble of random networks. The values of $k$ at which this \emph{normalized rich club coefficient} peaks are of particular interest and are indicative of possible rich clubs.

\subsubsection*{Rich club module density}
A rich club analysis specifies whether a node is part of a rich club at a particular $k$. From this binary assignment, we can ask how frequently rich club nodes are assigned to the same module. This measure, \emph{rich club module density}, is calculated as:

\begin{equation}
d_r = \frac{1}{|r|^2} \sum_{ij \in r} T_{ij}.
\end{equation}

\noindent In short, $d_r$ measures the average association weight between all pairs of rich club nodes.

\subsubsection*{Functional fingerprints}
Previous studies of brain functional connectivity networks -- the statistical similarity of brain regions' activity -- have shown that they can be partitioned into sub-systems that, broadly, are associated with one or more cognitive domains as determined by functional neuroimaging \cite{power2011functional,yeo2011organization,bellec2010multi}. Applying modularity maximization to anatomical networks usually yields modules that do not overlap exactly with the boundaries of these functional systems. To measure the extent to which any detected module, $c$, overlaps with a functional system, $s$, we calculated the Jaccard index:

\begin{equation}
J_{cs} = \frac{|c \cap s|}{|c \cup s|}.
\end{equation}

The numerator counts the number of regions that are jointly assigned to $c$ and $s$ while the denominator counts the number of regions assigned to $c$ or $s$. The value of $J_{cs}$ can be biased by the sizes of $c$ and $s$, so we standardized it against the null distribution obtained by randomly permuting module assignments 10000 times, and expressed the overlap as a $z$-score. We compared detected structural modules against the functional systems reported in \cite{mivsic2015cooperative}, which included: subcortical (SUB), temporal (TEMP), visual (VIS), somatomotor (SMN), dorsal attention (DAN), default mode (DMN), salience (SAL), control (CONT), and ventral attention (VAN) networks (Fig.~S7). A list of region-to-system assignments is now included as a supplementary item (ROINames.txt).

\subsubsection*{Module consistency score}
Our primary focus was on identifying modules for representative, group-level connectivity matrices. However, we also applied modularity maximization to individual subjects. For each consensus module detected in the group-level matrix, we calculated a region-level consistency score that measured, on average, how consistently that module was detected at the level of individual subjects. For a group-level consensus module, $c$, the consistency score was calculated by, first, identifying in each subject the module, $c'$ that maximized $J_{cc'}$. This yielded 30 modules -- one for each subject. For a node $i$ and consensus module $c$, we defined the consistency score as the fraction of those 30 modules, $c'$, in which node $i$ appeared.

\section*{Results}

\begin{figure*}[t]
	\begin{center}
		\centerline{\includegraphics[width=1\textwidth]{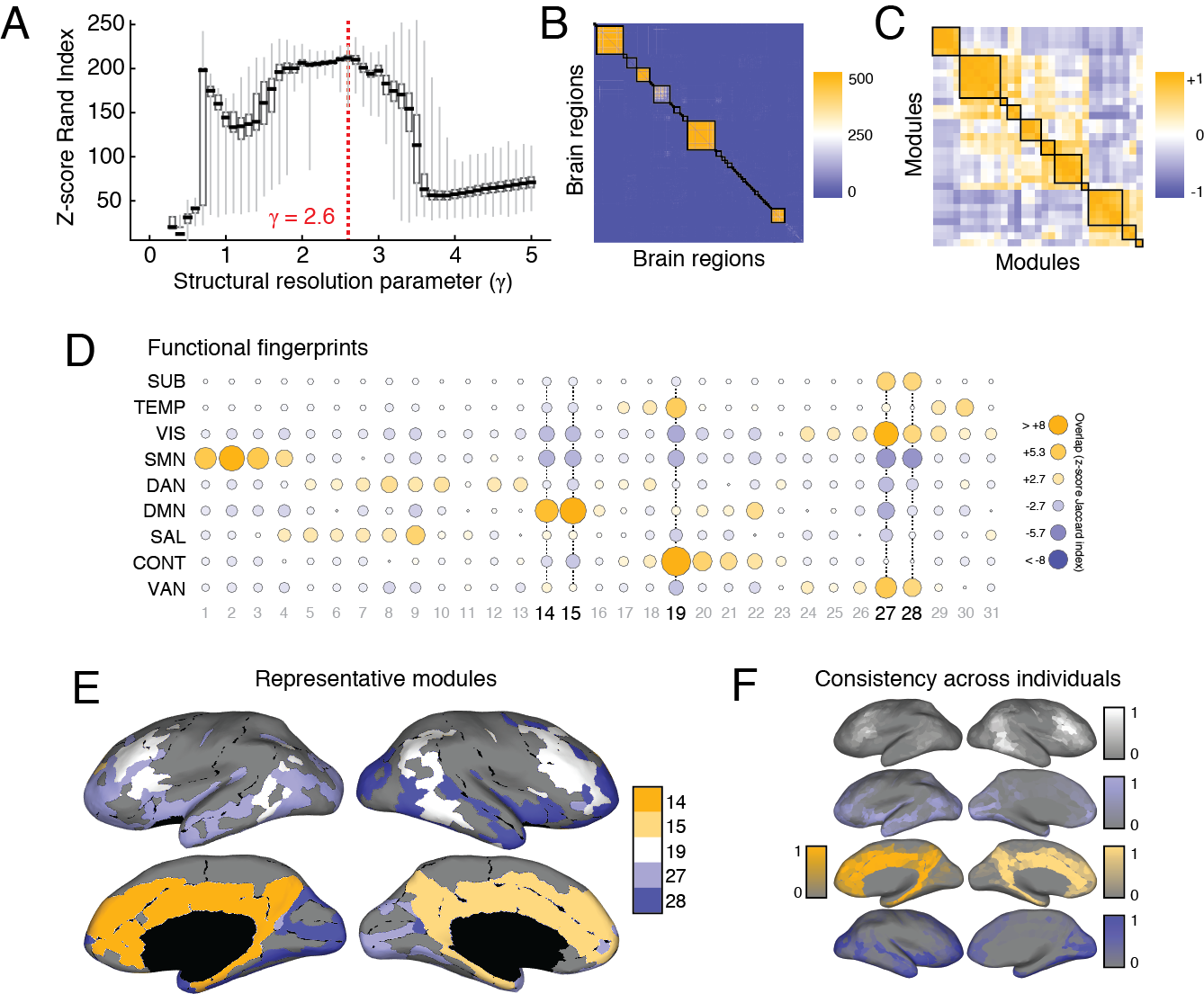}}
		\caption{\textbf{SPTL modules in Human DSI} (\emph{A}) Distribution of \emph{z}-score Rand indices as a function of $\gamma$. (\emph{B}) Association matrix (fraction of times out of 500 Louvain runs that each pair of nodes were assigned to the same module) clustered according to consensus modules. The 31 statistically signicant consensus modules exhibited correlated (\emph{C}) functional fingerprints (\emph{D}). (\emph{E}) Here we show the five largest consensus modules on the cortical surface. Each color corresponds to a different module. (\emph{F}) To demonstrate that these consensus modules, which we uncovered from a representative connectivity matrix, were also expressed at the level of individual subjects, we identified for each subject and for each consensus module, the module with greatest overlap and averaged the nodes that comprised that module to obtain a consistency score. The colorbars show the level of consistency across subjects.} \label{humanDSImodules}
	\end{center}
\end{figure*}

\subsection*{Characterizing modules detected using cost-reducing model}
In this report we maximized two different modularity functions in order to detect modules in human DSI and DTI datsets. The first modularity, $Q^{\text{NG}}$, compared the observed network with the standard Newman-Girvan (NG) null model. The second modularity, $Q^{\text{SPTL}}$, was novel and compared the observed network to a spatial (SPTL) null model tuned to match the brain's reduced wiring cost. Previous analyses of the brain's modular organization using the NG model have uncovered a small number of consistent, spatially-defined modules that overlap with functional systems \cite{hagmann2008mapping, bassett2010efficient, betzel2013multi}. The properties of modules detected using the SPTL model, however, are heretofore unknown. In this section, we characterize the topography, consistency, and functional fingerprints of these modules. Throughout the remainder of the paper, we refer to these null models as the SPTL and NG models and any modules detected using either model as SPTL or NG modules, respectively.

\subsubsection*{Human DSI}
We observed that the $z$-score of the Rand coefficient, a measure of partition similarity, achieved a local maximum at $\gamma = 2.6$ (Fig.~\ref{humanDSImodules}A), hinting at the presence of especially well-defined modules (see Methods). At that parameter value we uncovered a consensus partition of the brain into 82 modules, most of which were small (64 modules were made up of fewer than 10 brain regions). Of the modules detected at this scale, 31 were considered statistically significant ($p < 0.01$, corrected for false discovery rate) accounting for 731/1014 brain regions (Fig.~\ref{humanDSImodules}B). Many of the consensus modules spanned both hemispheres and exhibited non-random overlap with functional systems, which defines each module's \emph{functional profile} (Fig.~\ref{humanDSImodules}D). Moreover, modules' functional profiles were correlated with one another suggesting that the brain's long-distance modular architecture exists in a relatively low-dimensional space (Fig.~\ref{humanDSImodules}C). In Fig.~\ref{humanDSImodules}E we show a subset of five consensus modules. We focus on these modules because they were the largest and also because they were consistently expressed at the individual-subject level (Fig.~\ref{humanDSImodules}F). The first two modules, labeled 14 and 15, were bilaterally symmetric and spanned the medial surface. They included precuneus, components of anterior and posterior cingulate cortex, along with components of entorhinal, parahippocampal, and medial orbitofrontal cortex. Predictably, these modules exhibited the greatest overlap with the default mode network (DMN). Module 19 consisted of four spatially disjoint clusters spanning both hemispheres. It was composed, predominantly, of left and right inferior parietal and temporal cortex, middle frontal cortex, and \emph{pars opercularis}. The spatial topography of this module resembled the brain's control network (CONT). Finally, modules 27 and 28, which were also bilaterally symmetric, were situated inferiorly along the anterior-posterior axis. In addition to subcortical (SUB) structures caudate, putamen, pallidum, accumbens area, hippocampus, and amygdala, these modules were made up of regions in the visual (VIS) system, including lateral occipital, fusiform and lingual cortex. Additionally, these modules included regions from middle and orbito-frontal cortex as well as insular and temporal cortices, which mapped onto components of the ventral attention network (VAN). We show the smaller remaining modules in the supplement (Fig.~S8).

In the supplement we demonstrate the robustness of these consensus module assignments to variation in network node definition (\emph{Supplement -- Including versus excluding subcortical regions}; Fig.~S1),  resolution parameter value (\emph{Supplement -- Robustness to choice of resolution parameter}; Figs.~S2, S3), and tractography and network reconstruction parameters (\emph{Supplement -- Robustness to variation in max curvature angle}; Fig.~S4).

\begin{figure*}[t]
	\begin{center}
		\centerline{\includegraphics[width=1\textwidth]{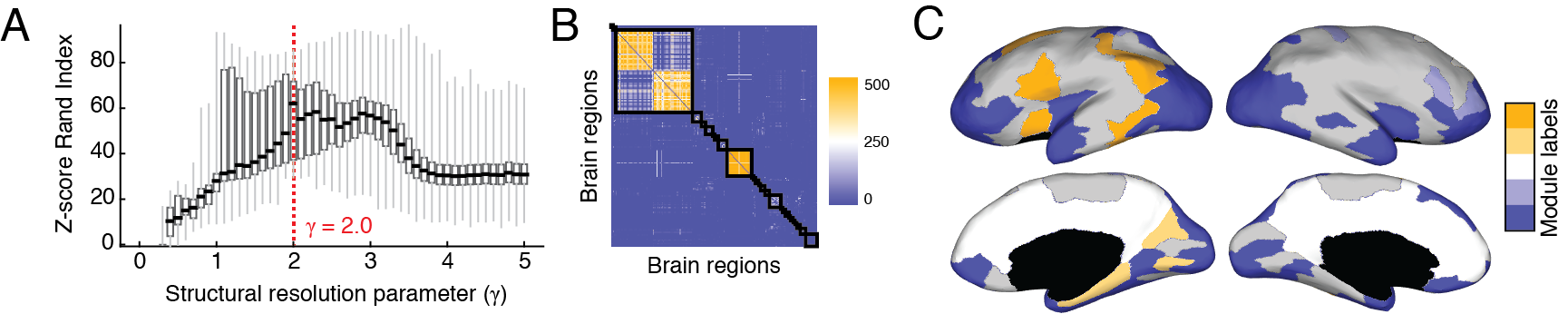}}
		\caption{\textbf{SPTL modules in Human DTI} (\emph{A}) Distribution of \emph{z}-score Rand coefficients as a function of $\gamma$. (\emph{B}) Association matrix (fraction of times out of 500 Louvain runs that each pair of nodes were assigned to the same module) clustered according to consensus modules. (\emph{C}) The five statistically significant consensus modules shown on the cortical surface. Each color corresponds to a different consensus module.} \label{humanPNCmodules}
	\end{center}
\end{figure*}

\subsubsection*{Human DTI}
We performed a similar analysis of the human DTI dataset. We observed a peak $z$-score Rand coefficient at $\gamma = 2.0$ (Fig.~\ref{humanPNCmodules}A). At this scale, we detected 39 modules, five of which were considered statistically significant (Fig.~\ref{humanPNCmodules}B). These modules accounted for 137/233 brain regions. While the statistically significant modules differed slightly from those detected in the human DSI connectome, they nonetheless had many features in common. The largest of the five modules (87 regions) largely recapitulated the inferior, bilateral modules (labeled 27 and 28 in Fig.~\ref{humanDSImodules}E), combining them into a single module. Indeed, upon examination of the association matrix, there was evidence suggesting an alternative consensus partition in which this single module gets split into two bilaterally symmetric modules (Fig.~\ref{humanPNCmodules}) The second-largest module (28 regions) similarly combined modules 14 and 15 into a single module. The remaining three modules accounted for 22 regions and resembled, albeit imperfectly, module 19. As a group, these final three statistically significant modules spanned both hemispheres.

\begin{figure*}[t]
	\begin{center}
 		\centerline{\includegraphics[width=0.8\textwidth]{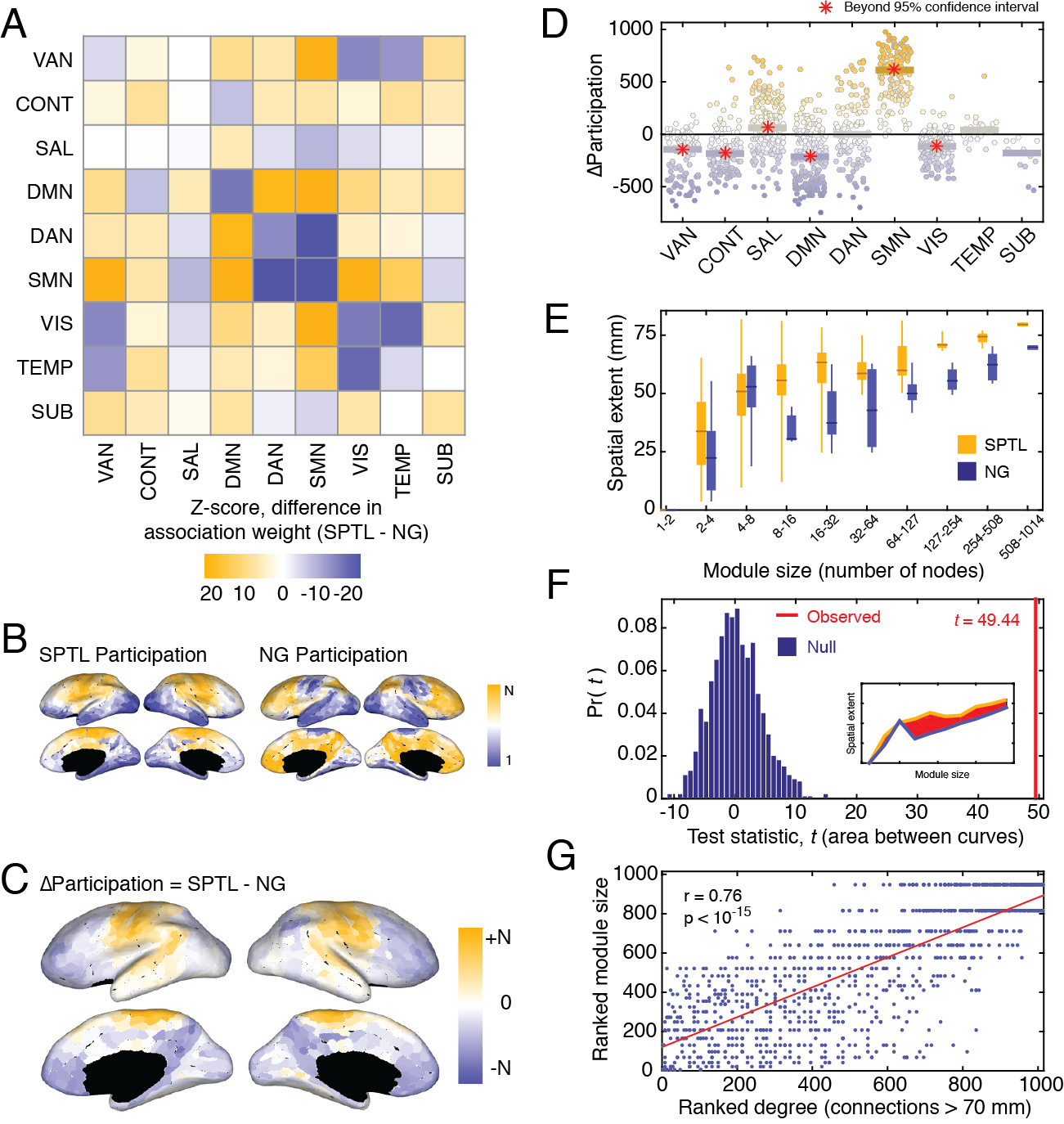}}
		\caption{\textbf{Comparing properties of SPTL and NG modules} (\emph{A}) Differences in SPTL and NG association matrices, grouped by functional systems and $z$-scored against a null distribution. (\emph{B}) Ranked node-level participation coefficients on the cortical surface based on the SPTL (\emph{left}) and NG (\emph{right}) modules. (\emph{C}) Difference between ranked SPTL and NG participation coefficients. (\emph{D}) System-level changes in participation coefficient. Each point represents a single brain region. The bars represent the median change in participation coefficient over all regions assigned to each system. A red star indicates that the median change exceeds the 95\% confidence interval of the null distribution. (\emph{E}) The spatial extent (mean interregional distance) of modules as a function of module size for both the SPTL and NG models. (\emph{F}) The area between the two module size versus spatial extent curves (\emph{inset}) serves as a test statistic under functional data analysis (FDA). We compared the observed statistic to test statistics estimated had module assignments been random. (\emph{G}) Correlation of a region's number of long-distance connections against the size of the module to which it was assigned.} \label{compareModels}
	\end{center}
\end{figure*}

\subsection*{Comparing SPTL and NG modules}
To better contextualize the SPTL modules, we contrasted them with NG modules. The NG model, when applied to the DSI data, exhibited a maximum $z$-score Rand coefficient at $\gamma = 1$, which resulted in a partition into four modules of $273$, $193$, $268$, and $280$ nodes (Fig.~S9). We also observed a second local maximum at $\gamma = 2.1$, which resulted in a finer partition of the network into $18$ smaller modules, including eight singletons. However, to maintain an analysis pipeline consistent with our investigation of the spatial null model, we focused on the division into four modules.

\subsection*{Changes in module association}
One of the most intuitive means of comparing SPTL and NG modules is to test whether, under one model or the other, certain pairs of nodes are more likely to be coassigned to the same module. To identify such pairings, we first subtracted the NG association matrix from the SPTL association matrix. The elements of the resulting matrix were positive or negative when node pairs were more likely to be coassigned to the same module under the SPTL or NG model, respectively. To further facilitate interpretation we aggregated these differences by functional systems and standardized these scores against null distributions obtained by randomly permuting system assignments (10000 permutations). Thus, for every pair of functional systems, we were left with a $z$-score indicating how much more likely it was for nodes in those systems to be coassigned to same module under SPTL model compared to the NG model (Fig.~\ref{compareModels}A). 

We observed that among functional systems, the somatomotor network (SMN) exhibited some of the most dramatic differences. Under the NG model, the SMN regions tended to be assigned to the same module as other SMN regions and components of the dorsal attention network (DAN) ($p <= 10 \times 10^{-15}$). Under the SPTL model, however, SMN regions were much more likely to appear in modules alongside the ventral attention network (VAN) and default mode network (DMN) ($p <= 10 \times 10^{-15}$). The DMN, itself, exhibited a distinct pattern. Whereas DMN regions tended to appear in the same module as one another under the NG model, they were more likely to appear in modules with all other systems under SPTL model (other than the control (CONT) network). Collectively, these results indicate that SPTL and NG modules exhibit different patterns of module co-assignment. An important question, then, is how these different patterns reshape our understanding of brain function.

\subsubsection*{Changes in participation coefficient}
Given a modular partition of a network, one can calculate the node-level metric \emph{participation coefficient}, which quantifies the extent to which a node's links are confined to its own module \emph{versus} spread out over different modules \cite{guimera2005functional}. A brain region's participation coefficient can be used to assess its integrative capacity -- i.e. whether or not that node links modules to one another \cite{hagmann2008mapping}. We calculated the participation coefficients for both SPTL and NG modules. Because the average participation coefficient is correlated with the size and number of modules in a partition and because we wished to compare partitions that differed in terms of these quantities, we rank-transformed the raw participation coefficients (Fig.~\ref{compareModels}B) before calculating the region-wise difference (Fig.~\ref{compareModels}C). To quantify which systems exhibited the biggest change in participation, we grouped regions by system, calculated the median change in participation over nodes assigned to each system, and compared that value to a null distribution obtained by permuting system assignments (Fig.~\ref{compareModels}D). We observed that salience (SAL) and somatomotor networks (SMN) exhibited statistically significant increases in their participation coefficients (median scores in excess of the 95\% confidence interval of the null distribution). We observed corresponding decreases in participation in ventral attention (VAN), control (CONT), default mode (DMN), and visual (VIS) networks (median scores less than the 95\% confidence interval). The temporal (TEMP) and subcortical (SUB) systems exhibited no changes. Collectively, these results suggest that by maximizing $Q^{\text{SPTL}}$ the salience and somatomotor systems appear to occupy, potentially, more integrative roles in the network by distributing a greater proportion of their connections across different modules. Conversely, the systems whose participation decreased can be thought of as becoming more autonomous and less integrated with the network as a whole.

In addition to comparing the participation coefficients obtained from the SPTL model with those obtained from the NG models, we also compared participation coefficients obtained from the NG model with those obtained from the previously-described functional partition \cite{mivsic2015cooperative}. This comparison was performed using precisely the same methods and resulted in a similar outcome; notably, that the somatomotor network exhibited increased participation compared to the other systems, which tended to decrease or stay the same (see \emph{Participation coefficients of structural versus functional partitions} and Fig.~S5).

\subsubsection*{Mean interregional distance}
One of the simplest statistics to compute over modules is the spatial extent of each module or the mean interregional distance among all nodes assigned to the same module \cite{muldoon2013spatially}. A module's spatial extent will tend to increase with its size, so we only compared spatial extents between similarly-sized communities. We observed that spatial extent increased more or less monotonically as a function of module size for both SPTL and NG modules. In other words, small modules tended to be made up of nearby nodes and, as modules grew in terms of number of nodes, they also tended to grow in terms of their spatial extent. However, for a given-sized module, the spatial extent of SPTL modules exceeded that of NG modules (Fig.~\ref{compareModels}E). As a means of quantifying this observation, we used \emph{functional data analysis} \cite{ramsay2002applied, ramsay2006functional}, which is a set of statistical tools for comparing continuous curves and has been previously used to study brain networks \cite{bassett2012altered}. Here, we defined two curves: the median interregional distance of modules as a function of module size, which we computed for both the SPTL and NG models. At each bin, we summed the difference between curves and compared this total difference (49.44) to what we would expect by chance (obtained from 1000 random permutations of module labels). In all cases, the observed difference was greater than random ($p \approx 0)$ (Fig.~\ref{compareModels}F), indicating that SPTL modules have broader spatial extent than NG modules and may, therefore, be driven more by costly long-distance connections than short-range, low-cost connections.

\subsubsection*{Singleton modules}
There are many brain regions that make few, if any, long-distance connections. For these regions, the SPTL model (especially for larger values of $\gamma$) might anticipate all of their existing connections. Accordingly, no grouping of these regions into a module can lead to an increase in modularity. This leads to a large number of small (or even singleton) modules. Indeed, at $\gamma = 2.6$, of the 82 modules, eleven were singletons and 64 were comprised of less than 11 nodes ($\approx 1\%$ of the total number of network nodes). Accordingly, we hypothesized that brain regions that make fewer long-distance connections will tend to be associated with smaller modules and \emph{vice versa}. To test this hypothesis, we calculated each node's distance-dependent degree -- i.e. its total number of connections greater than a certain distance. For each distance threshold, we calculated the correlation of this value with the size of the consensus module to which it was assigned. Indeed, at a distance threshold of 70 mm we found $r \approx 0.76$ ($p < 10^{-15}$) (Fig.~\ref{compareModels}G). This suggests that one of the principal drivers of module size is the number of long-distance connections that a node makes.

\begin{figure*}[t]
	\begin{center}
		\centerline{\includegraphics[width=1\textwidth]{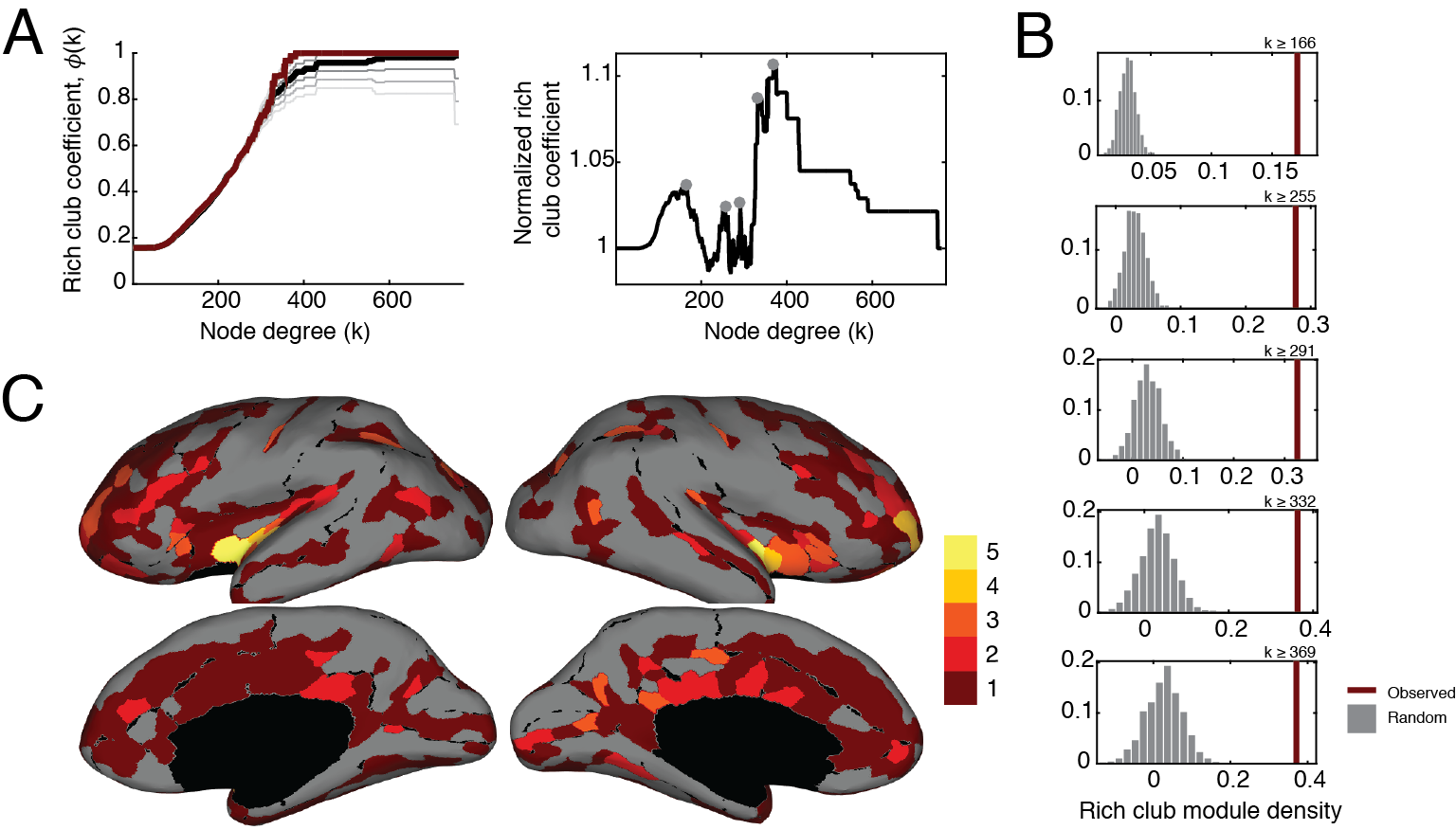}}
		\caption{\textbf{Results of rich club analysis} (\emph{A}) (\emph{left}) Raw rich club coefficient (in red) as a function of node degree. The black line represents the mean rich club coefficient over 1000 random (degree-preserving) networks. Each increasingly brighter gray line represents one standard deviation away from that mean (up to three standard deviations). (\emph{right}) The normalized rich club coefficient (observed raw divided by random mean), which exhibited five local maxima at $k = 166, 255, 291, 332, 369$. (\emph{B}) Rich club module density, $d_r$, for each of the five rich clubs we investigated. In red is the observed value of $d_r$, while the gray shows the null distribution of densities over 10000 randomizations. (\emph{C}) Topographic visualization of rich club consistency across the cerebral cortex (note: sub-cortical regions are not pictured). Colored regions indicate how many of the five rich clubs a region participated in, with brighter colors indicating greater participation.} \label{richClubComparison}
	\end{center}
\end{figure*}

\subsection*{Relationship to rich clubs}
The rich club phenomenon -- the propensity for high-degree nodes to be more densely interconnected than expected -- is ubiquitous in biological neural networks. The current interpretation of the rich club is as an integrative structure, with spatially-distributed rich-club nodes linked by costly long-distance connections serving as bridges from one module to another and acting as a backbone over which information from one module can be rapidly transmitted to another module \cite{van2012high,van2013anatomical}. Most papers discussing the relationship of rich clubs to modules have used modularity maximization in conjunction with the NG model. This leads to two important observations: (\emph{1}) the rich club is never detected as a cohesive module (although blockmodels may prove useful in this endeavor \cite{pavlovic2014stochastic}) and (\emph{2}) the interpretation of the rich club as an integrative structure, in part, depends upon how modules are defined. Accordingly, we wished to compare the relationship of SPTL and NG modules to rich clubs. To facilitate such a comparison, we first calculated the normalized rich-club coefficient, which exhibited several distinct peaks, suggesting the existence of multiple rich clubs of different sizes. We focused on five of these peaks, which corresponded to rich clubs of brain regions with $k \ge$ 166, 255, 291, 332, and 369 (the corresponding sizes of rich clubs were 402, 98, 46, 20, and 14 regions) (Fig.~\ref{richClubComparison}A,B). In addition to sub-cortical regions (which were part of the rich club at all scales), we observed that bilateral insula, rostral middle frontal, superior parietal, and superior temporal cortex were consistently assigned to the rich club, in agreement with previous studies \cite{van2011rich,van2013anatomical}.

Rich club regions tend to be linked by long connections, but the modules detected using the NG model have short spatial extents. This makes it unlikely that rich club regions will be co-assigned to the same module. Modules detected using the SPTL model, on the other hand, have broader spatial extent, meaning that they potentially could co-assign many rich club regions to the same module. To test for this possibility, we calculated the average rich club module density for both the SPTL and NG models, across all values of $\gamma$, and for each of the five rich clubs. We observed that the rich club module density was consistently greater than expected for the SPTL model compared to the NG model (Fig.~\ref{richClubComparison}C). This result suggests that the modules detected using the SPTL null model better recapitulate relationships among rich club nodes than those detected using the NG model.

\subsection*{Space-independent modules across development}
\begin{figure*}[t]
	\begin{center}
		\centerline{\includegraphics[width=1\textwidth]{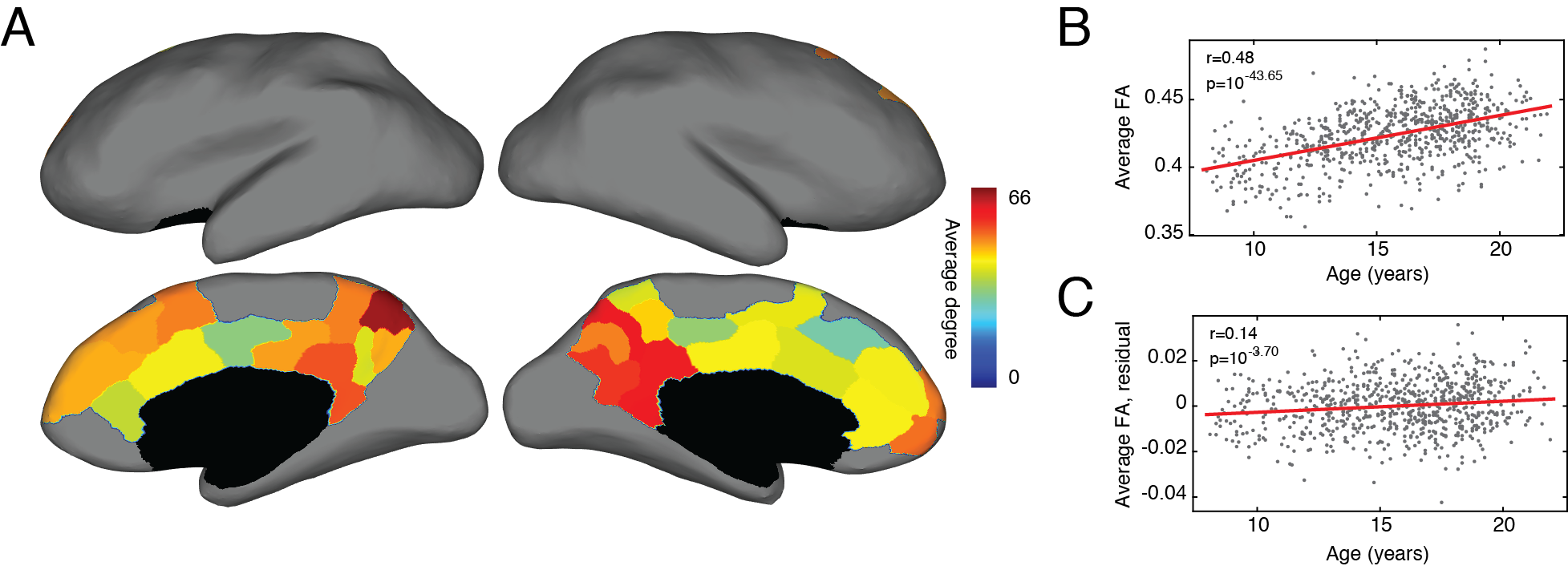}}
		\caption{\textbf{SPTL modules over development} (\emph{A}) We observed that a single module exhibited changes in its average internal white matter integrity (as measured by fractional anisotropy; FA). This module consisted of precuneus, posterior cingulate, and anterior cingulate cortex. The color of each region indicates its degree in the representative connectivity matrix (\emph{B}) Without correction for confounding variables, we observe a statistically significant increase in the average within-module FA over development. (\emph{C}) This relationship, albeit attenuated, persists after correcting for confounding network, physiological, and morphological variables.} \label{development}
	\end{center}
\end{figure*}

Finally, we used the modules we detected in the Human DTI dataset to highlight changes in development. Over normative development, the brain refines its white- and gray-matter \cite{gogtay2004dynamic}, and the underlying anatomical network becomes increasingly similar to the pattern of functional couplings \cite{hagmann2010white}. Concurrently, brain development is paralleled by profound intellectual and cognitive growth \cite{casey2000structural}, suggesting that the two processes may be interrelated. Here, we assessed whether SPTL modules tracked development. Specifically, we calculated the average within-module fractional anisotropy (FA), a measure of fiber integrity, and asked whether this variable was correlated with a participant's age. We found that, before accounting for confounding variables, twelve modules exhibited statistically significant age-related changes ($p<0.01$, FDR-corrected). The strongest correlation was for a bilateral midline module comprised of precuneus, posterior cingulate, and anterior cingulate cortex (Fig.~\ref{development}A), whose within-module FA was correlated with age (Pearson correlation coefficient of $r = 0.48$, $p < 10^{-15}$; Fig.~\ref{development}B).

FA and other network statistics, however, can be influenced by a number of confounding variables. For example, head motion induces systematic biases in FC measurements \cite{satterthwaite2012impact, power2012spurious}. Similarly, normal changes in brain and intracranial volume over the lifespan can serve as a source of unwanted variation \cite{betzel2014changes}. How these variables influence structural connectivity measures and what processing strategies might reduce their influence are not well understood. Nonetheless, we sought to minimize the influence of confounding variables by regressing them out of the within-module FA scores and calculating the correlation of the residuals with age. In all, we identified binary density (total number of connections), global FA (averaged over all connections), signal to noise ratio, an estimate of head motion, and total intracranial volume as potential confounds. Regressing out these variables lead to a reduction in the number of modules whose within-module FA was correlated with age from twelve down to one; the lone surviving community was the midline module ($r = 0.14$, $p < 2^{-4}$) (Fig.~\ref{development}C). This result suggests that normative development can, in part, be characterized by refinement of white matter connections within a single module that includes precuneus and both anterior and posterior cingulate cortex.

\section*{Discussion}
Modularity maximization with the Newman-Girvan (NG) null model represents the standard module detection method in network neuroscience. This standard persists despite technical and philosophical issues with the NG model. In this report, we suggest that a cost-reducing model in which longer connections are formed with decreasing probability represents an appropriate null model for modularity maximization, as it dually mimics the brain's preference for short low-cost connections while allowing us to investigate in a principled way the organization of long-distance, costly connections. Using this SPTL model, we show that the detected modules diverge from those we detect using the NG model and exhibit distinct functional fingerprints. We also show that, as measured by changes in the participation coefficient, the somatomotor network appears as a more integrative structure while default mode, control, ventral attention, and visual systems appear more autonomous and segregated. Additionally, the SPTL model yields modules that are more similar to the brain's rich clubs as compared to the NG model. Finally, we show that, when applied to a developmental cohort, we uncover a module situated bilaterally along the midline whose internal density of connection weights increases significantly with age. Collectively, these results support the hypothesis that the brain exhibits consistent, non-random, and spatially-defined modules. Unlike previous reports, the modules we uncover cannot so easily be explained on the basis of a cost-reduction principle, suggesting that they may be of added functional importance. Collectively, these results complement our current conception of the brain's modular organization and offer additional insight into its functional roles.

\subsection*{Why a cost-reduction null model?}
The main goal of this paper was to shift focus away from modules driven by short-range connections and onto modules driven by unexpected, costly, and long-distance connections. Besides this goal, the use of a cost-reduction null model confers other distinct advantages. First, as noted in recent work in condensed matter physics, a good null model should have many features in common with the observed network \cite{bassett2015extraction}. While there is no definitive list of which features should and should not be preserved, the conclusions one can draw will depend on whether the null networks are physically viable \emph{versus} not physically viable (c.f. \cite{bassett2015extraction} \emph{versus} \cite{bassett2012influence}). These considerations are likely particularly important for spatial networks, like those considered here, where each edge is associated with a cost; the NG model, because it allows for the formation of long-distance connections with no penalty, gives rise to exceedingly costly null networks. This fact motivates an exploration into the effects of alternative null models, both on spatial structure and -- in the future -- on system dynamics \cite{papadopoulos2016evolution}.

A second reason for considering a cost-reduction model, specifically, is because it effectively shifts the focus from short connections on to long connections. Why might this be advantageous? As noted earlier, long connections are costly and require more energy to sustain compared to short connections. Additionally, long connections are also costly in terms of their volume (the total volume of all connections needs to fit within the skull \cite{sherbondy2009think}) and their computational capacity (long connections imply longer processing delays \cite{cuntz2010one}). Therefore, over evolutionary time, we might expect that such costly features would fade away; either the brain regions linked by costly connections would grow closer (in space) so that the same connection is effectively shortened, or whatever functions the long connection supports would be taken over by different regions linked by shorter connections. The fact that we still observe long, costly connections means that they likely perform specific functions that cannot easily be performed by other regions. By comparing our network against a cost-reduction null model, we effectively make long connections the focal point of our analyses, which may help us better understand their functional roles more precisely. 

A third and final point for considering a cost-reduction model is that in reality we have no ``ground truth'' knowledge about the brain's modular organization. It is unclear whether the best description of the brain's modules comes from a block model \cite{pavlovic2014stochastic} or whether the modules should be allowed to overlap \cite{de2014edge}. Moreover, even in networks where the ground truth modules are known, module detection techniques tend to perform poorly. For example, in annotated social networks where an individual's affiliation with a particular social group can be determined unambiguously, many module detection techniques fail to detect these ground truth groups based on connectivity alone \cite{hric2014community, yang2014structure}. Accordingly, over-interpreting the output of any single module detection algorithm (or null model) may therefore be ill-advised. A more balanced approach would be to compare results of multiple methods to achieve greater intuitions for the architectures that support the brain's complex dynamic repertoire.

\subsection*{What does the SPTL model tell us about brain function?}

Applying graph theory to the connectome helps us generate hypotheses about how a brain functions as a network. The consensus point of view is that the connectome's main function is to regulate and constrain brain dynamics \cite{deco2011emerging}, structuring the flow of information from one brain region to another \cite{van2012high,goni2014resting}. Different network attributes are thought to contribute in different ways. Modules, for instance, are thought to be useful for local, segregated information processing, while ``shortcuts'' and hubs are viewed as integrative structures for rapidly transmitting information over long distances. With these intuitions in mind, one can make predictions about individual brain regions' functional roles based on how they are situated within the network. One such measure is the participation coefficient, which considers how a region's links are distributed across modules \cite{guimera2005functional}. Regions whose links are distributed across different modules (participation coefficients close to one) are thought to help regulate intermodular communication, whereas regions with low participation (close to zero) might play a greater role in effecting communication patterns within their own module. 

The consensus has been that regions along the midline -- e.g. precuneus, posterior cingulate, anterior cingulate -- are among the brain's hubs -- they tend to have high degree and also high participation (in some cases, both) and (perhaps unsurprisingly) are believed to play important roles in intermodular communication \cite{hagmann2008mapping}. These same regions are also considered parts of the default mode, salience, control, and attention networks \cite{power2011functional, yeo2011organization}, suggesting that higher-order cognitive systems might owe part of their functionality to the fact that their components span multiple modules and can efficiently integrate information from those sources \cite{van2013anatomical,bertolero2016modular}. Additionally, these regions are also among the most vulnerable in psychiatric and neurodegenerative diseases, such as schizophrenia and Alzheimer's \cite{van2013network}. However, participation coefficient is always defined with respect to a particular set of modules. Here, we demonstrated that nodes' participation coefficients exhibit stereotypical differences when we define modules using the NG model compared to the SPTL model. In particular, we find that regions within the somatomotor network are uniquely positioned to have high participation, suggesting a greater integrative (though not necessarily influential) role for that network. The opposite is true for higher-order functional systems that saw their participation coefficients decrease. These differences can be used in the future to better understand structural constraints on cognitive flexibility, cognitive control, and attention \cite{gu2015controllability,medaglia2016cognitive}.

\subsection*{Space-independent modules across development}
In the final component of this report, we uncovered a module made up of precuneus and both posterior and anterior cingulate cortex. We observed that the average FA of fiber tracts within this module increased significantly with age, even after controlling for confounding variables. The composition of this module is of particular interest, as it overlaps closely with both the medial components of the default mode network \cite{raichle2001default,raichle2015brain,andrewshanna2010functional} and with putative rich club regions \cite{van2011rich}.

Because FA is often interpreted as a measure of fiber integrity, this result suggests the maturation of these structures over the course of development. In general, this result agrees with previous developmental studies of the brain's maturing structural architecture, which have revealed that increases in FA edge weights contribute to an increased correspondence of structural and functional connections \cite{hagmann2010white, supekar2010development}. Our finding also agree with observations that the rich club is already well-defined in children but undergoes subtle changes across development to reach its mature state \cite{grayson2014structural}.

\subsection*{The merits of modularity maximization}

In this report we have pointed to a number of drawbacks to applying modularity maximization in conjunction with the NG model to discover modules in human connectome data. Despite this, modularity maximization as a general method remains one of the most commonly used techniques in network science, broadly, and for good reason. Indeed, there are many reasons for using modularity maximization. First, there are many fast heuristics for maximizing a modularity quality function. These include spectral methods \cite{newman2006modularity}, greedy algorithms \cite{blondel2008fast}, and belief propagation \cite{zhang2014scalable}, to name a few. Additionally, certain heuristics for maximizing modularity \emph{can} lead to highly accurate results when applied to networks with planted structural modules \cite{lancichinetti2009community,bassett2013robust}, suggesting that under a certain set of assumptions modularity maximization can be expected to deliver good results. Finally, modularity maximization as a general framework is readily extended to multi-slice networks \cite{mucha2010community} and can accommodate a multitude of different null models \cite{expert2011uncovering, traag2011narrow, aldecoa2013exploring,bassett2015extraction,papadopoulos2016evolution, nicolini2016modular}. Collectively, this set of properties -- easily implemented, highly accurate under some circumstances, and highly flexible -- make modularity maximization a reasonable option for performing module detection. Here, we simply demonstrate that informing the modularity quality function with spatially grounded null models may be an important direction for future research.

\subsection*{Methodological considerations}
There are a number of methodological considerations to take into account, both in our approach to detecting network modules, but also in terms of how we interpret them once they are detected.

\subsubsection*{Why not apply modularity maximization to reduced-cost networks?}
In this report, we extended the modularity maximization framework by changing what it means for a connection to be expected. Specifically, we selected a null model (the SPTL model) where connection formation depended on the brain's spatial embedding, being tuned to match the brain's preference for short-range connections. Another possibility, and one that has been explored previously, is to use the same SPTL to produce an ensemble of graphs, maximize the modularity of each graph in the ensemble using the NG model, and compare the resulting modules to those observed by applying modularity maximization to the real brain network \cite{henderson2011geometric, henderson2013using, samu2014influence, roberts2015contribution}. In general, these methods share the view that the observed network modules are quite similar to those obtained with the other null models; this observation is perhaps not so surprising considering that the brain is comprised predominantly of short-range connections, so the cost-reduction model that \emph{also} features many short-range connections ought to have not dissimilar modular structure. Our findings, while buttressed by these earlier studies, are distinct, however. By discounting short-range connections, we make it possible to detect entirely novel module organization, whereas the earlier analyses could only confirm that two sets of modules were similar to one another.

\subsubsection*{Interpretation of modules}
Modularity maximization seeks to identify modules -- collections of nodes that are more densely connected to one another than expected by chance. Here, we compare modules detected with the NG null model, which is the standard in the field, with those obtained when we use a SPTL null model. We show that, in most cases, the modules that we obtain with the SPTL model are unique in both their composition (the nodes that belong to that module) and their topography (the distribution of nodes across the brain). This occurs because the change in null models shifts the algorithm's focus from modules that are simply denser than expected to modules that have more long distance connections than expected. It is worth noting, however, that even with this shift in focus it is in principle possible to detect precisely the same modules with both null models. This could occur if a module satisfies both conditions.

\subsection*{Limitations of modularity maximization}
Here, we took advantage of the generic nature of the modularity maximization framework to define a novel modularity function in which we compared observed brain networks with a null connectivity model based on wiring reduction principles. While modularity maximization is flexible to alternative null models \cite{expert2011uncovering} and has proven useful in detecting modules in multi-layer networks \cite{mucha2010community}, it has a number of important shortcomings. First, for certain classes of null models (including the Newman-Girvan model) modularity exhibits a so-called resolution limit \cite{fortunato2007resolution}, in which it is incapable of resolving communities smaller than a characteristic scale. While the inclusion of a resolution parameter makes it possible to shift this scale and thereby detect smaller modules, it does not fully mitigate the effects of the resolution limit \cite{lancichinetti2011limits}. In addition, and as we noted earlier, modularity maximization is also prone to a degeneracy of near optimal partitions \cite{good2010performance}. We attempted to deal with this problem by focusing on consensus modules rather than any single estimate of modules.

Another potential issue involves our use of a consensus clustering approach for resolving the variability of the modules detected over multiple runs of the Louvain algorithm. This procedure involved using modularity maximization to recluster partitions estimated using modularity maximization. While consensus clustering leads to more accurate estimates of a network's community structure \cite{lancichinetti2012consensus} and the self-consistency of the current implementation is appealing, it also means that any biases exhibited by modularity maximization or the Louvain algorithm are doubly present.

Despite its shortcomings, modularity maximization remains a flexible method for identifying a network's modules. This is due, in part, to the fact that the modularity equation can accommodate alternative models of null connectivity -- a fact that we take advantage of in the present study. While it is generally considered good practice to verify that modules detected using one algorithm are, at the very least, qualitatively similar to those detected using another \cite{hric2014community,peel2016ground}, we are unaware of other module detection algorithms that are readily capable of accepting alternative null connectivity models. Accordingly, we were unable to verify the robustness of the modules we detected using different algorithms. Future work will be directed towards the development of alternative methods for detecting network modules while controlling for spatial relationships.

\subsubsection*{Extensions}
In this report we analyzed binary networks, meaning that connections between brain regions have weights of one (if a link was detected) or zero (if no link was detected). While this binary link structure is of importance -- the presence or absence of a link certainly acts to constrain communication patterns between brain regions -- by discarding information about the relative strength of a link, which could be encoded with a real-valued weight, we throw away some information about the network's organization and function. In principle, at least, our model can be extended to the case of weighted networks -- we could replace the probability distribution for the presence/absence of edges to consider their weights, as well. This is a slightly more complicated model, and we do not explore it here. Additionally, a weighted and signed variant of this model would make it possible to apply a similar method to functional connectivity networks, which are often defined as correlation matrices.

\subsubsection*{Use of spatially-embedded null models for more general comparisons of brain networks}
Our study leveraged a cost-reduction model for exploring the brain's modular structure. It serves as a convenient foil in that it explicitly tries to account for properties of the network that are driven by space and cost-reduction principle. This approach can be (and in some cases has been) used more generally to test whether space influences other network properties, for example the propensity for regions to form a rich club or the distribution of hub regions across the brain \cite{samu2014influence, roberts2015contribution}. For example, putative rich clubs are identified by comparing an observed rich club coefficient against that of a null model \cite{van2011rich}. As with modularity maximization, this model is typically selected to be the NG model. It may be the case that comparison against a different model could reveal rich clubs of different compositions than what we typically observe.

\subsubsection*{Diffusion imaging and tractography}
We construct brain networks from diffusion imaging and tractography data, both of which have notable advantages but also drawbacks. Presently, these methods represent the state-of-the-art (and only) techniques for the non-invasive reconstruction of structural brain networks \cite{wandell2016clarifying}. Despite this, it has been shown that tractography may be insensitive to white matter tracts that run parallel to the cortical surface \cite{reveley2015superficial} and that tractography algorithms may be prone to algorithm-specific biases \cite{thomas2014anatomical}. Nonetheless, under ideal circumstances, diffusion imaging and tractography can do reasonable jobs reconstructing known anatomical tracts \cite{calabrese2015diffusion} and, with the advent of new algorithms and techniques, will surely show improvement as the field matures \cite{pestilli2014evaluation, merlet2012tractography}.

Interestingly, the SPTL model may be useful for correcting biases in the tractography algorithms themselves. In this report, we frame the formation of short-range connections as being driven by a cost-reduction mechanism. Short-range connections, however, could also appear due to biases in tractography that make it easier to track short, within-hemisphere connections compared to long, inter-hemispheric connections \cite{girard2014towards}. In principle, then, the SPTL model's parameters could be tuned to match the characteristics of short-range false positive connections, thereby allowing us to focus more clearly on non-artifactual connections.

\section*{Conclusion}
In conclusion, our work expands on previous studies showing that much of brain network architecture can be attributed to a cost-reduction principle. We go one step further and search for features of the network that are inconsistent with such a principle. We reveal a novel set of modules that, because they cannot be accounted for by a cost-reduction principle, may be of particular functional significance. We show that these exhibit distinct properties and change with normative development.

\bibliography{spatial_brain_modules_biblio}

\clearpage
\beginsupplement

\section*{Contents of this Supplementary Document}

This file includes the following content:
\begin{enumerate}
	\item Supplementary Materials and Methods
	\item Figures S1 to S9
	\item References
\end{enumerate}

\section{Supplementary Materials and Methods}
In the main text we presented an analysis of the modular organization of human structural brain networks. Specifically, we proposed a modification of the well-known modularity quality function \cite{newman2006modularity}, wherein we replace the null connectivity model with one that depends upon the brain's spatial embedding. In this supplement we present a number of additional analyses that demonstrate the robustness of our results.

\subsection{Including \emph{versus} excluding subcortical regions}
In the main text we focused on a network composed of $N = 1014$ brain regions. These regions were based on a subdivision of the so-called Desikan-Killany atlas \cite{desikan2006automated}. That atlas consists of 68 cortical regions and 14 subcortical regions. The subdivision was constructed by dividing the 68 cortical regions into 1000 approximately equal volume regions \cite{cammoun2012mapping} while not sub-dividing subcortical regions at all. As a consequence, the 1000 cortical regions tend to be of smaller volume than the 14 subcortical regions. In general, regions with larger volume will tend to have higher degree which can bias the topology of the resulting network. To avoid these biases a number of recent studies using the same parcellation have opted to exclude all sub-cortical regions from analysis \cite{goni2014resting, betzel2013multi, mivsic2015cooperative, avena2014using}. Here, we demonstrate that the optimal consensus modules we describe in the main text are relatively robust to our decision of whether to include or exclude subcortical regions in the network.

\begin{figure*}[t]
	\begin{center}
		\centerline{\includegraphics[width=1\textwidth]{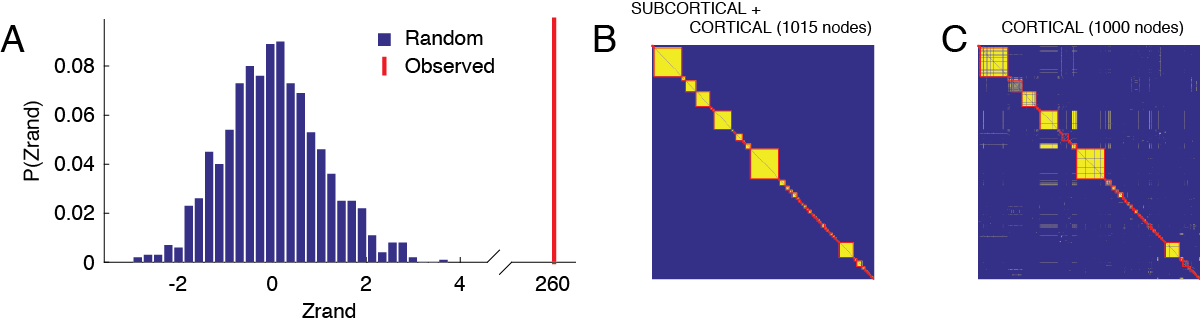}}
		\caption{\textbf{Comparison of communities with/without subcortical regions}. To test whether the inclusion of subcortical regions influenced detected modules, we constructed a group-representative network comprised of $N=1000$ cortical regions and performed modularity maximization using SPTL model on this network. (\emph{A}) We found that the optimal cortical partition was detected at $\gamma = 2.8$. Comparing the optimal cortical and subcortical + cortical partitions, we observed that they were highly similar to one another ($z_{rand} = 260.26$; maximum value of null distribution $z_{rand} = 3.40$, 10000 permutations). (\emph{B}) Subcortical + cortical parcellation association matrix ordered by optimal subcortical consensus partition. (\emph{C}) Cortical parcellation association matrix ordered by optimal subcortical consensus partition.} \label{subcortex_vs_neocortex}
	\end{center}
\end{figure*}

In parallel to our analysis presented in the main text, we generated and analyzed a group-representative connectivity matrix comprised of only the $N=1000$ cortical regions. To differentiate this matrix from the one in the main text, we refer to the network analyzed in the main text as SUBCORTICAL + CORTICAL and the cortical only network as CORTICAL. Our analysis of the CORTICAL network was performed in precisely the same manner as our analysis of the SUBCORTICAL + CORTICAL network -- e.g. we selected the optimal resolution parameter value based on the median partition similarity and we constructed a consensus partition at that parameter value. For the CORTICAL network we observed an optimal resolution parameter of $\gamma = 2.8$ (the SUBCORTICAL + CORTICAL network described in the main text achieved its optimal resolution parameter at $\gamma = 2.6$). To assess the similarity of the CORTICAL and SUBCORTICAL + CORTICAL consensus partitions, we compared them using the z-score of the Rand coefficient \cite{traud2011comparing}. Specifically, we removed the sub-cortical regions from the SUBCORTICAL + CORTICAL partition so that both partitions contained the same 1000 nodes, calculated their similarity, and compared this value to a null distribution obtained by randomly and uniformly permuting the module assignments of the CORTICAL partition 10000 times. The observed similarity of the two partitions was $z_{rand} = 260.26$ while the largest largest similarity in the null distribution was $z_{rand} = 3.40$ (Fig.~\ref{subcortex_vs_neocortex}A). This suggests that the two partitions are similar to each other well beyond what would be expected under this specific null model. We support this statistical analysis qualitatively by showing both CORTICAL and SUBCORTICAL + CORTICAL association matrices side by side with their rows and columns ordered such that nodes assigned to the same module in the SUBCORTICAL + CORTICAL partition were positioned next to eachother (Fig.~\ref{subcortex_vs_neocortex}B,C).

\subsection{Robustness to choice of resolution parameter}
Multi-scale modularity maximization involves modifying the modularity quality function by including the resolution parameter, $\gamma$ \cite{reichardt2006statistical}. This modification makes it possible to tune $\gamma$ to different values and uncover communities of different sizes, thereby partially mitigating the effect of the so-called ``resolution limit'' \cite{fortunato2007resolution}. Unfortunately, there is no agreed upon method for selecting the optimal value of $\gamma$; most studies avoid the issue by setting its value to its default setting of  $\gamma = 1$ or by using some heuristic rule for choosing its value. In the main text we selected $\gamma$ to be the value at which the detected partitions were most similar to one another (as measured by the z-score of the Rand coefficient). This resulted in us focusing on communities detected at $\gamma = 2.6$.

\begin{figure*}[t]
	\begin{center}
		\centerline{\includegraphics[width=1\textwidth]{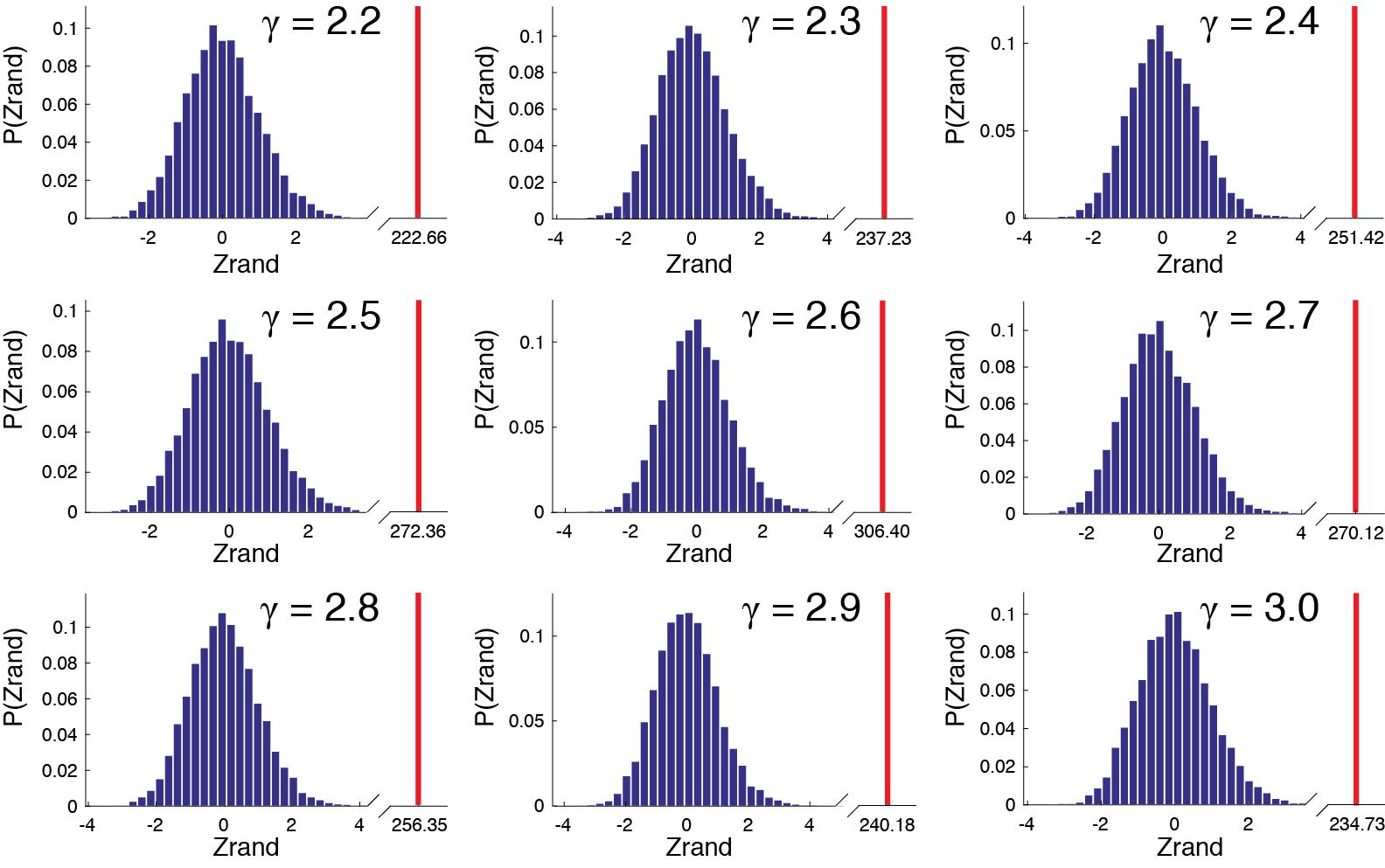}}
		\caption{\textbf{Statistical assessment of community robustness to variation in $\gamma$}. We tested how similar consensus partitions from $\gamma = 2.2$ to $\gamma = 3.0$ were to the partition obtained at $\gamma = 2.6$. We calculated the similarity (z-score Rand coefficient) of each consensus partition with respect to the partition obtained $\gamma = 2.6$ and then generated a null distribution by calculating a comparable similarity score but where the community labels in each consensus partition were uniformly permuted at random.} \label{robustnessOfPartitions}
	\end{center}
\end{figure*}

It is possible that different heuristics could highlight different resolution parameters. While it is beyond the scope of this paper to systematically compare such heuristics, we wish to demonstrate that our consensus modules are robust to reasonable variations in $\gamma$. Accordingly, we compared consensus partitions obtained at $\gamma = 2.2, \ldots , 3.0$ in increments of 0.1 (nine total values) with those described in the main text. As in the previous section, our comparison involved calculating the z-score of the Rand coefficent and comparing it against a null distribution obtained through permutation testing. In general, we found that all of the consensus partitions detected in this range were much more similar to those detected at $\gamma = 2.6$ than would be expected by under the null model (the minimum observed z-score over any of the nine $\gamma$ values was 222.66 while the maximum value obtained in any of the null distributions was 5.43) (Fig.~\ref{robustnessOfPartitions}).

\begin{figure*}[t]
	\begin{center}
		\centerline{\includegraphics[width=1\textwidth]{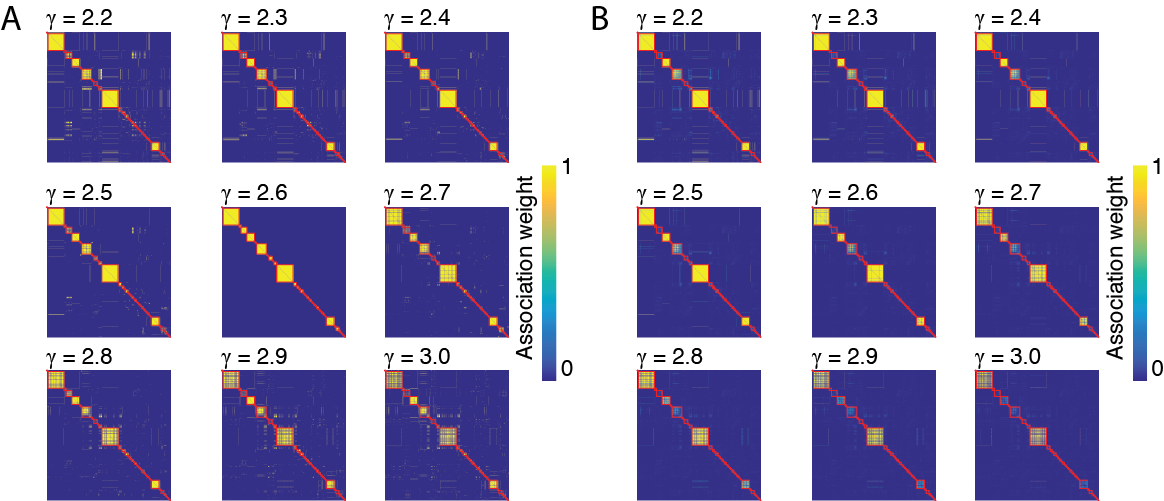}}
		\caption{\textbf{Qualitative assessment of community robustness to variation in $\gamma$}. As a qualitative assessment of similarity, we show association matrices over the same range of $\gamma$ constructed from consensus partitions and the ensemble of partitions obtained from the initial modularity maximization. (\emph{A}) Association matrices constructed from consensus partitions. (\emph{B}) Association matrices constructed from the initial modularity maximization.} \label{robustnessOfPartitions2}
	\end{center}
\end{figure*}

In addition to this statistical analysis, we show the association matrices calculated from the consensus modules detected over this same range. To facilitate comparision, we order their rows and columns according to the consensus partition obtained at $\gamma = 2.6$ (described in the main text). We see that, qualitatively, the partitions are excellent matches and that most non-zero elements are located within modules (the blocks along the diagonal) (Fig.~\ref{robustnessOfPartitions2}A). We also show, for the same values of $\gamma$, the association matrices calculated from the ``raw partitions'' (i.e. those obtained from modularity maximization but before consensus clustering). (Fig.~\ref{robustnessOfPartitions2}B).

\subsection{Robustness to variation in max curvature angle}
Reconstructing connectivity matrices from diffusion imaging data with tractography algorithms is challenging and prone to both false positives and negatives \cite{thomas2014anatomical, reveley2015superficial}. In addition, tractography algorithms include multiple parameters that can influence the resulting matrices. One important parameter is the max curvature angle that determines the largest orientation change that a streamline can exhibit between integration steps before it is terminated. While it is beyond the scope of the present study to systematically test all possible values of this parameter (and others), we nonetheless wanted to demonstrate that our results are consistent if we vary its value within some reasonable range.

The subject-level networks described in the main text were reconstructed using a max curvature angle of 35$^\circ$. Accordingly, we repeated our analysis using networks reconstructed with max curvature angles of 30$^\circ$ and 40$^\circ$. From these networks, we followed the methods described in the main text to construct group-represenative networks. As, perhaps, was expected, the group-representative networks varied in terms of gross network statistics. For example, the total number of connections increased monotonically with maximum curvature angle: 51994, 80513, and 119204 connections for the 30$^{\circ}$, 35$^{\circ}$, and 40$^{\circ}$ cutoffs, respectively.

\begin{figure*}[t]
	\begin{center}
		\centerline{\includegraphics[width=0.65\textwidth]{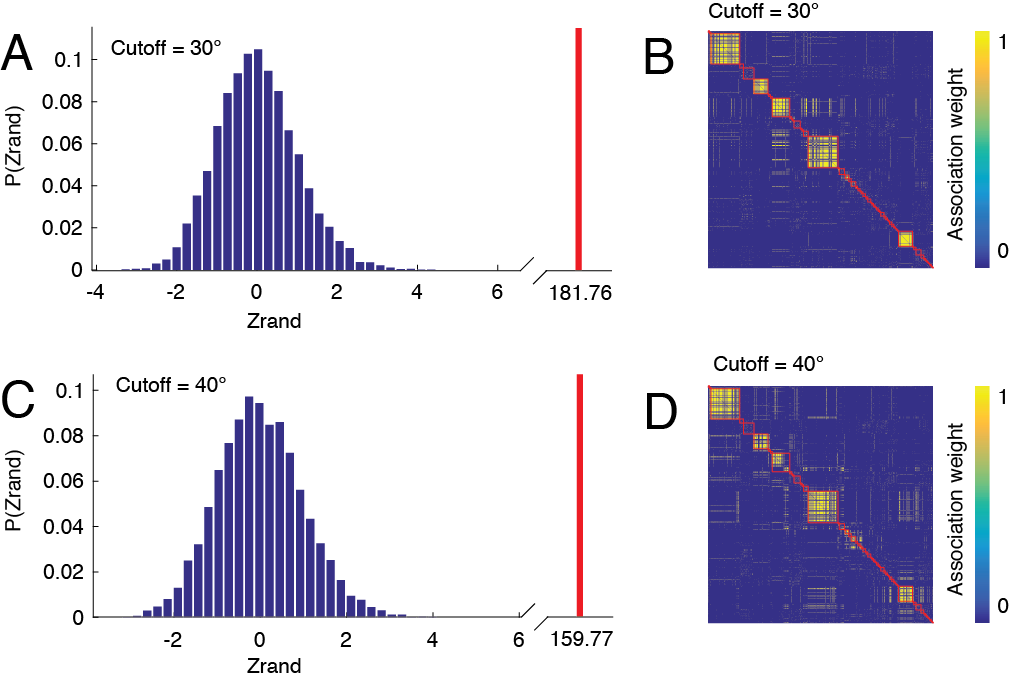}}
		\caption{\textbf{Comparison of consensus modules obtained using different maximum curvature cutoffs}. (\emph{A,C}) Z-score Rand coefficents of consensus modules described in main text with consensus modules obtained using different tractography parameters -- specifically maximum curvature angles of 30$^{\circ}$ and 40$^{\circ}$. We compare a null distribution (blue) versus what was observed (red), noting that in both cases the observed value exceeds the maximum values of the null distribution by a wide margin. (\emph{B,D}) Association matrices for consensus modules obtained using 30$^{\circ}$ and 40$^{\circ}$ maximum curvature angles. The rows and columns of each matrix have been ordered according to the consensus modules described in the main text. Note that most non-zero elements tend to fall within modules, indicating high similarity.} \label{otherCurvatures}
	\end{center}
\end{figure*}

Our analysis of these networks were identical to what was described in the main text. Specifcally, we maximized the SPTL modularity over a range of resolution parameters and identified the optimal parameter as the one that maximized the mean pairwise similarity of partitions detected at that value. We found that the for the 30$^{\circ}$ and 40$^{\circ}$ cutoffs, the optimal resolution parameters were $\gamma = 3.0$ and $\gamma = 2.1$, respectively. Once the optimal parameter was selected, we used the consensus clustering procedure to generate consensus modules.

We then calculated the similarity of these new consensus modules with the consensus modules described in the main text (where similarity was measuresd as the z-score Rand coefficient). To contextualize these scores, we compared them against a null distribution obtained by randomly and uniformly permuting module assignments. The observed similarity of the consensus modules obtained with maximum curvature angles of 30$^{\circ}$ and 40$^{\circ}$ to the consensus modules described in the main text were 181.76 and 159.77, respectively, which far exceeded the maximum values obtained in either null distribution (4.48 and 4.28) (Fig.~\ref{otherCurvatures}A,C). Based on these observations, we conclude that the consensus modules obtained using 30$^{\circ}$ and 40$^{\circ}$ cutoffs are, at least in this specific statistical sense, similar to the those described in the main text. For qualitative assurance that the modules were similar to one another, we also generated association matrices for the consensus modules obtained using 30$^{\circ}$ and 40$^{\circ}$ cutoffs and ordered their rows and columns according to the module assignments described in the main text (obtained using 35$^{\circ}$ cutoff). Our expectation was that most non-zero elements would fall within modules, whereas few non-zero elements would appear between modules. Indeed, this is the case (Fig.~\ref{otherCurvatures}B,D)

\subsection{Participation coefficients of structural \emph{versus} functional partitions}

\begin{figure*}[t]
	\begin{center}
		\centerline{\includegraphics[width=0.75\textwidth]{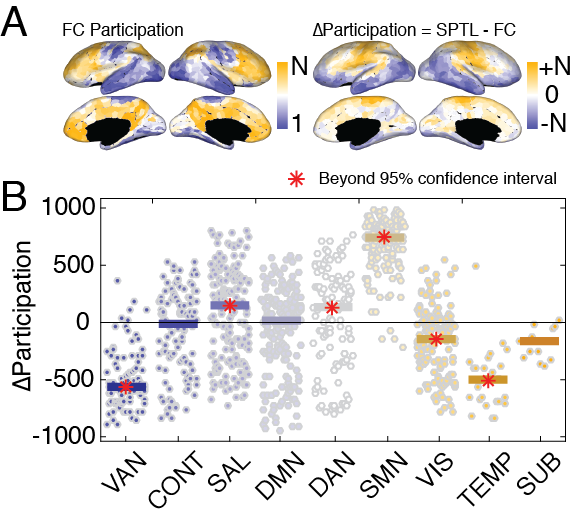}}
		\caption{\textbf{Comparison of participation coefficients}. (\emph{A}) Ranked participation coefficients obtained using functional connectivity (FC) partition (from \cite{mivsic2015cooperative}). (\emph{B}) Difference in ranked participation coefficient obtained from SPTL model and FC partition. (\emph{C}) Regional participation coefficients grouped by functional system.} \label{fcParticipation}
	\end{center}
\end{figure*}

In the main text we described brain regions' participation coefficients with respect to \emph{structural partitions} obtained from structural connectivity networks. Specifically, we compared structural partitions obtained by maximizing the standard Newman-Girvan modularity with those obtained using a novel space-dependent modularity.  This approach -- hub classificiation based on structural partitions -- has been described in a number of high profile studies and we sought to continue this tradition \cite{sporns2007identification, hagmann2008mapping}. However, a number of other studies have calculated participation coefficients with respect to a \emph{functional partition} obtained, for example, from an ICA analysis of fMRI BOLD time series \cite{van2013anatomical} or from community structure analysis of functional connectivity brain networks \cite{warren2014network}. This second approach makes it possible to assess indirectly the roles of individual nodes with respect to the brain's functional systems.

Therefore, in the interest of completeness, we compared participation coefficients calculated using the consensus partition obtained from the SPTL model with those obtained using the functional partition described in the main text (taken from \cite{mivsic2015cooperative}). We followed precisely the same methods as described in the main text: (1) we ranked participation coefficients, (2) subtracted one set of coefficients from the other, (3) and grouped these coefficients by functional system to estimate the mean change difference in participation coefficient for each system (the statistical significanc of which we assessed using a permutation test). We also produced surface plots to show the spatial distribution of participation coefficients obtained using the functional partition. Though the results of this procedure are not in precise agreement with those described in the main text, they are qualitatively quite similar serve to highlight the same changes in participation. Notably, both comparisons highlight the somatomotor network as exhibiting increased participation.

\section{Additional Supplementary Figures}

\begin{figure*}[t]
	\begin{center}
		\centerline{\includegraphics[width=0.5\textwidth]{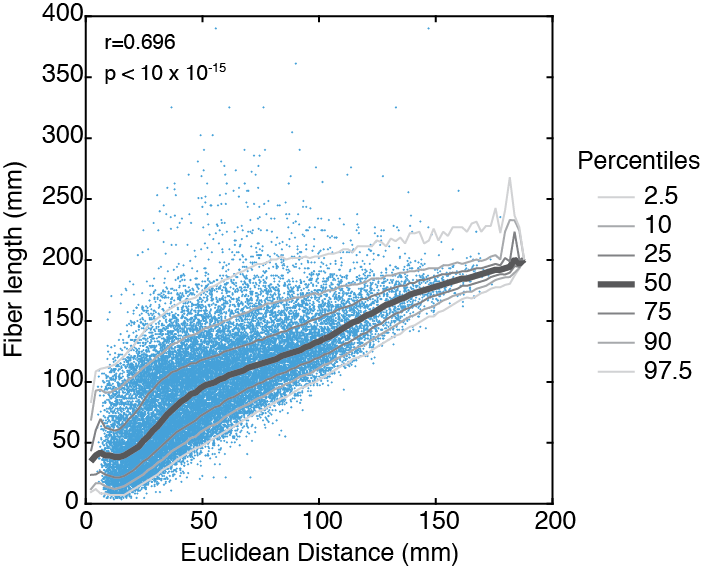}}
		\caption{\textbf{Correlation of fiber length and Euclidean distance} We show a strong correlation of a connection's Euclidean distance (straight line distance between its endpoints) and its fiber length (curvilinear  trajectory through space).} \label{fiberLength}
	\end{center}
\end{figure*}

\begin{figure*}[t]
	\begin{center}
		\centerline{\includegraphics[width=1\textwidth]{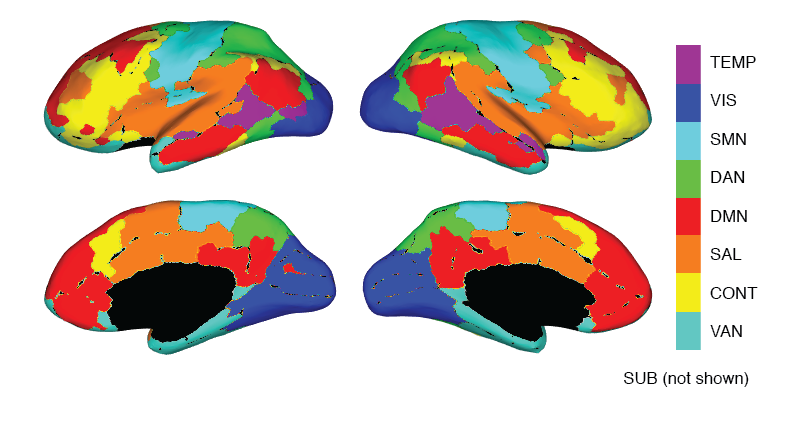}}
		\caption{\textbf{Functional systems} Topographic representation of 8/9 functional systems; the subcortical system (SUB) is not shown in this rendering. The systems shown are: temporal (TEMP), visual (VIS), somatomotor (SMN), dorsal attention (DAN), default mode (DMN), salience (SAL), control (CONT), and ventral attention (VAN) networks.} \label{functionalSystems}
	\end{center}
\end{figure*}

\begin{figure*}[t]
	\begin{center}
		\centerline{\includegraphics[width=1\textwidth]{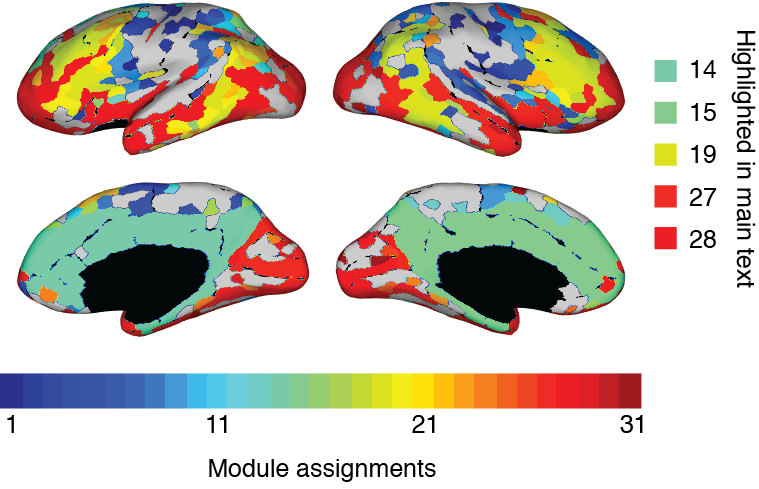}}
		\caption{\textbf{All detected consensus SPTL modules for DSI dataset} Topographic representation of the 31 consensus modules detected by applying modularity maximization using the SPTL null model to the human DSI datset.} \label{HumanDSIOtherCommunities}
	\end{center}
\end{figure*}

\begin{figure*}[t]
	\begin{center}
		\centerline{\includegraphics[width=1\textwidth]{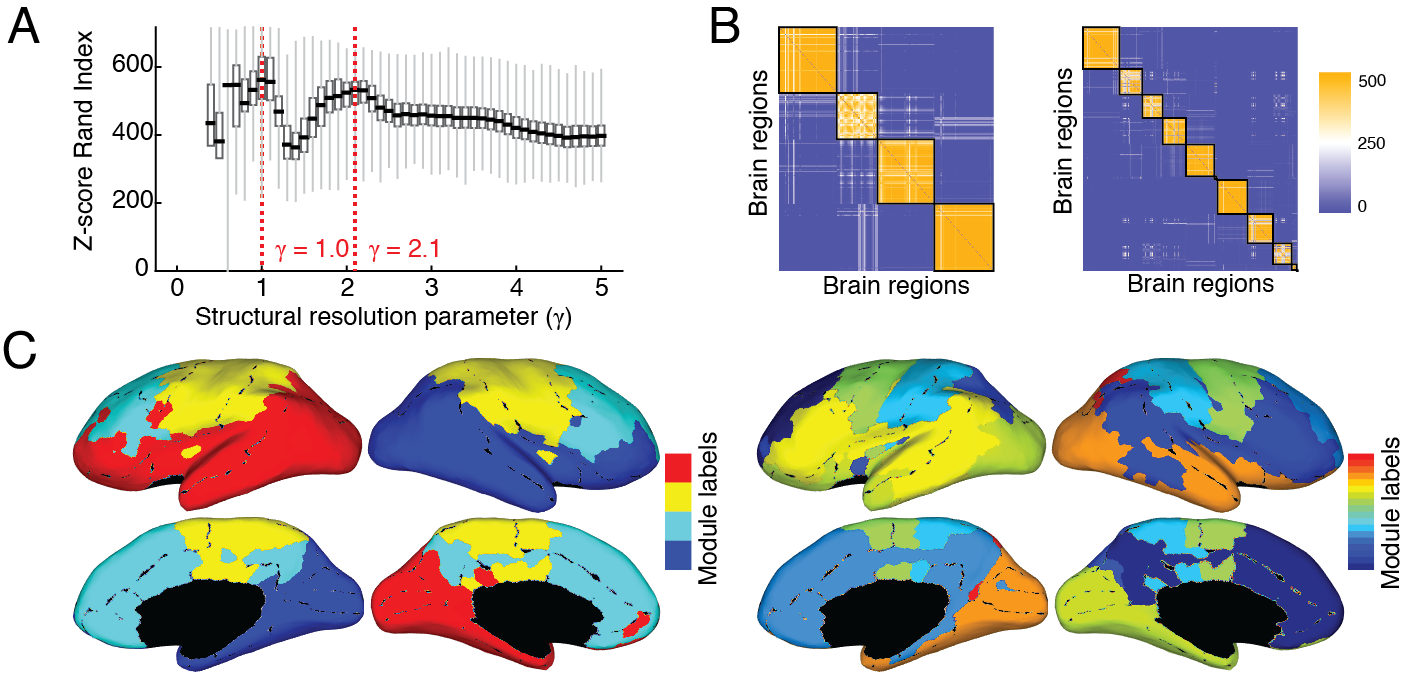}}
		\caption{\textbf{NG modules in Human DSI} (\emph{A}) Distribution of \emph{z}-score Rand indices as a function of $\gamma$. (\emph{B}) Association matrices (fraction of times out of 500 Louvain runs that each pair of nodes were assigned to the same module) clustered according to consensus modules for the two peaks in the curve shown in panel \emph{A}, corresponding to $\gamma = 1.0$ (\emph{left}) and $\gamma = 2.1$ (\emph{right}). (\emph{C}) Module assignments on the cortical surface for modules detected at $\gamma = 1.0$ (\emph{left}) and $\gamma = 2.1$ (\emph{right}).} \label{humanDSImodules_NG}
	\end{center}
\end{figure*}

\clearpage

\bibliography{spatial_brain_modules_biblio}

\end{document}